\documentclass[AMA,STIX1COL]{WileyNJD-v2}
\usepackage{graphicx,  setspace,  amsmath,  amssymb,  subfigure,  url,  multirow,  verbatim,  algpseudocode,  bm, threeparttable, color,  colortbl, booktabs, longtable, multirow, graphicx, float}
\usepackage{amsmath}
\usepackage{arydshln}
\usepackage{amsthm,  rotating}

\articletype{RESEARCH ARTICLE}%

\begin{document}

\title{Robust structured heterogeneity analysis approach for high-dimensional data}

\author[1,2]{Yifan Sun*}

\author[2]{Ziye Luo*}

\author[1,2]{Xinyan Fan}

\authormark{SUN \textsc{et al}}

\address[1]{\orgdiv{Center for Applied Statistics}, \orgname{Renmin University of China}, \orgaddress{\state{Beijing}, \country{China}}}

\address[2]{\orgdiv{School of Statistics}, \orgname{Renmin University of China}, \orgaddress{\state{Beijing}, \country{China}}}

\corres{Xinyan Fan, No. 59 Zhongguancun Street, Haidian District, Beijing, 100872. \\
\email{1031820039@qq.com}}

\abstract[Summary]{Revealing relationships between genes and disease phenotypes is a critical problem in biomedical studies. This problem has been challenged by the heterogeneity of diseases. Patients of a perceived same disease may form multiple subgroups, and different subgroups have distinct sets of important genes. It is hence imperative to discover the latent subgroups and reveal the subgroup-specific important genes. Some heterogeneity analysis methods have been proposed in recent literature. Despite considerable successes, most of the existing studies are still limited as they cannot accommodate data contamination and ignore the interconnections among genes. Aiming at these shortages, we develop a robust structured heterogeneity analysis approach to identify subgroups, select important genes as well as estimate their effects on the phenotype of interest. Possible data contamination is accommodated by employing the Huber loss function. A sparse overlapping group lasso penalty is imposed to conduct regularization estimation and gene identification, while taking into account the possibly overlapping cluster structure of genes. This approach takes an iterative strategy in the similar spirit of K-means clustering. Simulations demonstrate that the proposed approach outperforms alternatives in revealing the heterogeneity and selecting important genes for each subgroup. The analysis of Cancer Cell Line Encyclopedia data leads to biologically meaningful findings with improved prediction and grouping stability.}

\keywords{Subgroup identification, High-dimensional data, Robustness, Overlapping clusters}

\maketitle

\footnotetext{*equal contributions from these authors.}

\section{Introduction}
Detection of relationships between the genes and the disease outcomes or phenotypes is one of the core tasks in biomedical research. 
Regression models are extensively used to link genes to a phenotype of interest and predict patients' phenotypic traits. Traditionally, a unified model is built with the assumption that all patients follow the same regression model. However, the heterogeneity of disease challenges the rationality and correctness of this practice. In fact, common diseases including cancer are highly heterogeneous. It has been widely acknowledged that various subtypes exist for these diseases, which vary in underlying etiology and pathogenesis \cite{Curtis2012,Wang2020Epigenetically,Samra2019Resolving}. As a result, patients of different subtypes may have different sets of important genes and varied genetic effects on phenotype. This fact indicates that instead of building a unified regression model for all patients, we should construct a set of heterogeneous regression models, each of which corresponds to one subgroup of patients. 

The key challenge of constructing a heterogeneous regression model is that the set of patients in each subgroup is not known \emph{a prior}. To tackle this challenge, multiple approaches have been proposed. The first one is the finite mixture regression (FMR), which is perhaps the most popular approach in heterogeneity analysis \cite{Mclachlan2000Finite}. Coupled with regularization and other techniques, FMR can also accommodate high-dimensional data \cite{Khalili2007Variable,Khalili2013Regularization,Lloyd2018}. The second one is penalized fusion, which discovers the latent subgroups by penalizing the pairwise differences of regression coefficients \cite{Ma2016A, Ma2020Exploration,Hu2021}. Recently, the penalized fusion has been generalized to the high-dimensional regression setting by combining with a sparse penalization and model averaging strategy \cite{He2020}.    

\begin{figure}
	\centering  
	\subfigure
	{
   	  \includegraphics[width=0.42\textwidth]{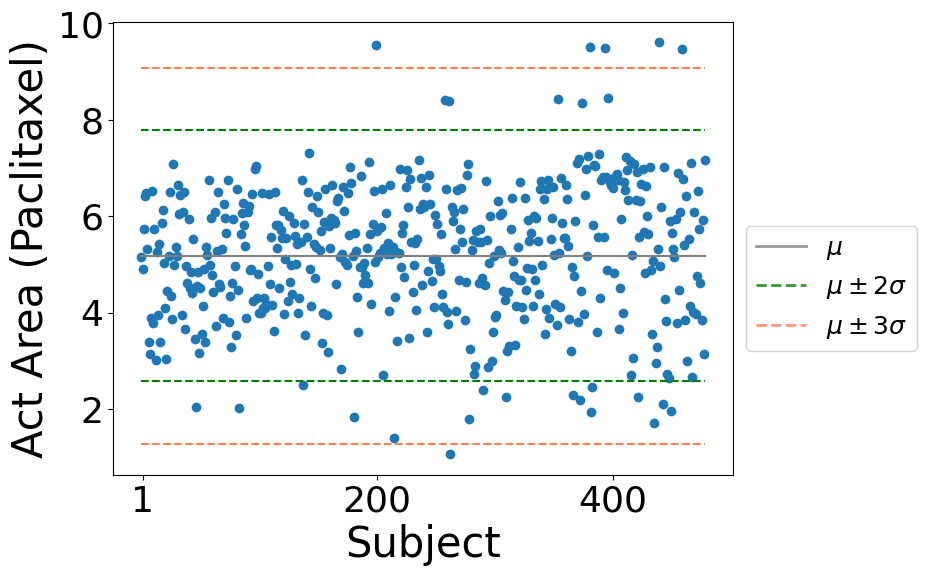}
	}
	\subfigure
	{
	 \includegraphics[width=0.55\textwidth]{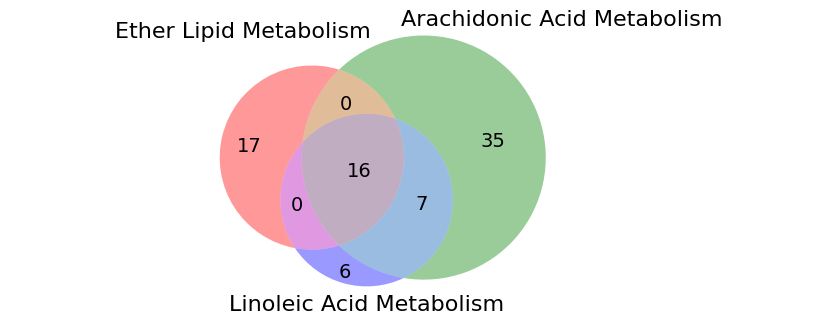} 
	}
	\caption{Analysis of CCLE data. Left: scatter plot of sensitivity score for the drug Paclitaxel. Right: Venn diagram of genes for three pathways (Ether Lipid Metabolism, Arachidonic Acid Metabolism, and Linoleic Acid Metabolism). The numbers show the size of overlap among pathways. }
	\label{fig:y_pacli}
\end{figure}

Despite considerable successes, the existing approaches for heterogeneity analysis may still be limited in the following aspects. First, most of the existing studies conduct standard least-square-based estimation, and thus cannot accommodate outliers/contamination in the response variable. In biomedical research, responses with outliers/contamination are often encountered, and can be caused by errors in data recording, biased sample selection, and other factors \cite{Osborne2004The}. To fix the idea, consider the Cancer Cell Line Encyclopedia (CCLE) data analyzed in our study (please refer to Section \ref{sec:ccle} for more details). This dataset entails 24 drug compound screening data for 947 cancer cell lines with associated gene expression and cell lines' sensitivity scores to all drugs. In the left panel of Figure \ref{fig:y_pacli}, we present the scatterplot of the sensitivity score (response variable) for Paclitaxel, a novel antimicrotubule drug used for multiple cancer treatments, as well as mean and three times standard deviation. It can be observed that six samples are out of the three times standard deviation range, suggesting that the data have some outliers with respect to response. Nonrobust methods fail to take robustness into account and may lead to false identifications of subgroups and important genes, as well as biased estimation of genetic effects. A number of robust analysis methods have been developed for high-dimensional data under the homogeneous regression models, and they have significant advantages over nonrobust methods for noisy data. For the heterogeneous regression models, there are a few robust analysis methods based on modified FMR \cite{Yao2014Robust,Song2014Robust,Hu2017The}, but the development is still much limited, especially for the high-dimensional data. 
The second limitation of the existing approaches is that there is insufficient attention to the interconnections among genes. Specifically, genes are organized into multiple known functional clusters, e.g., pathways. It is commonly recognized that many genes have several biological functions and thus belong to more than one functional cluster. For example, 16 genes, including JMJD7-PLA2G4B, PLA2G10, and PLA2G12A, have been suggested to participate in at least three pathways, i.e., Ether Lipid Metabolism, Arachidonic Acid Metabolism, and Linoleic Acid Metabolism (right panel of Figure \ref{fig:y_pacli}). Published studies have demonstrated the advantages of accommodating (possibly) overlapping cluster structure of genes in terms of prediction and gene identification \cite{Rao2016Classification}; however, these studies only focus on homogeneous regression models, and the studies in heterogeneous regression models are still lacking. 

Motivated by the aforementioned limitations in current studies, we propose a novel and efficient heterogeneity analysis approach in high-dimensional regression setting to identify the underlying subgroups in terms of gene-phenotype relationships, as well as select subgroup-specific important genes and estimate their effects. 
The proposed approach distinguishes itself from existing ones in the following aspects. First, the Huber loss function is adopted to accommodate outliers/contamination in response. Compared to alternative robust approaches, including the popular quantile regression-based approach, Huber approach has less computation cost, which is crucial for high-dimensional analysis. Second, a sparse overlapping group lasso penalty is introduced to accommodate the overlapping cluster structure of genes. Compared to the well-known group lasso penalty, it does no longer force all genes in a cluster to be selected and thus offers higher degree of flexibility in selecting important genes. Last but not the least, the proposed approach borrows the idea of K-means clustering method. Specifically, it aims to solve the minimization of a penalized objective function over possible groupings by using some iterative algorithms. Published studies show that the clustering-based method has many desirable advantages, including low requirements for data, implementation simplicity, low computation cost, and high scalability \cite{Bonhomme2015Grouped, Zhang2019Quantile}. However, the existing studies only focus on low-dimensional data, and cannot be directly applied to high-dimensional genomic data. To this end, sparse penalization in the form of a sparse overlapping group lasso is introduced to accommodate high dimensionality and distinguish important genes from noises. Our numerical study suggests favorable practical performance of the proposed approach. Overall, this study is warranted by providing a practically useful new venue for exploring the heterogeneity with respect to gene-phenotype relationships.

\section{Methods}
\subsection{Models and Estimators} 
 Consider a dataset with $n$ independent samples. For the $i$th sample, let $y_i$ be the response variable, representing the phenotypic trait, and $X_i=(X_{i1},\ldots,X_{ip})^{\top}$ be the $p$-dimensional vector of genes (gene expression, single-nucleotide polymorphisms, or other genetic functional units). We assume that the $n$ samples form $K\geq 2$ different subgroups, each of which is defined by a distinct relationship between the phenotype of interest and the genes, that is, 
\begin{equation}
\label{eq:model}
y_i=X_i^{\top}\beta_{g_i}+\varepsilon_i, 
\end{equation}
where $g_i$'s are the group memberships that take values in set $\{1,\ldots. K\}$, $\beta$'s are unknown regression coefficients, and $\varepsilon_i$'s are the random errors with an \emph{unknown} distribution function. Intercept is omitted for the simplicity of notation, and will be specified if necessary.
 Here we assume that the number of subgroup $K$ is known, and defer the discussion on how to determine $K$ in practice to Section \ref{sec:imp}. 
In our analysis, suppose the $p$ genes can be arranged into $L$ (possibly) \emph{overlapping} clusters $\mathcal{C}=\{C_1,\dots,C_L\}$ with $p_{l}$ genes in cluster $l$. The gene clusters are defined based on prior biological knowledge, .e.g, published information about biochemical pathways or co-expression in previous experiments.     

Our goal is threefold: one is to identify the group membership $g_i$, another is to select the important genes in each subgroup, and the other is to estimate their subgroup-specific coefficients $\beta_{g}$ for $g=1,\ldots,K$. Let $\mathcal{G}=\{g_1,\ldots,g_n\}$ and $\bm\beta=(\beta_1,\ldots,\beta_K)$, where $\beta_k = (\beta_{k1},...,\beta_{kp})^{\top}$ is the $p$-dimension regression coefficients for subgroup $k$. Thus, $\mathcal{G}\in \Gamma_{K}$ denotes a particular partition of the $n$ samples, where $\Gamma_K$ is the set of all partitions of $\{1,\ldots,n\}$ into $K$ subgroups. To achieve the aforementioned goal, we propose a novel estimate of $\bm\beta$ and $\mathcal{G}$, which is defined as the solution of the following minimization problem: 
\begin{equation}\label{eq:obj}
(\hat{\bm\beta},\hat{\mathcal{G}})=\mathop{\arg\min}_{\bm\beta\in\mathbb{R}^{p\times K}, \mathcal{G}\in\Gamma_K}  \mathcal{L}(\bm\beta, \mathcal{G})\equiv\sum_{k=1}^{K}\sum_{\{i|g_i=k\}}\rho_\delta(y_i - X_i^{\top} \beta_k) + \sum_{k=1}^K \phi(\beta_k; \gamma_k,\lambda_k),
\end{equation}
where the minimum is taken over all possible grouping $\mathcal{G}$ of the $n$ samples into $K$ subgroups and subgroup-specific regression coefficients $\bm\beta$. 

In the objective function $\mathcal{L}(\bm\beta, \mathcal{G})$ in Eq. (\ref{eq:obj}), the first term is the lack-of-fit. To accommodate outliers/contamination in the response, we adopt the Huber loss function, which takes the form 
\begin{equation}
\label{eq:loss}
	\rho_\delta(t) = \left\{
\begin{array}{lcl}
\frac{1}{2}t^2      &     & |t|\leq \delta \\
\delta |t|-\frac{1}{2}\delta^2      &     & |t|> \delta,
\end{array} \right.
\end{equation} 
where $\delta > 0$ is a tuning constant. Following published literature \cite{Friedman2001}, we set $\delta = 1.345\text{MAD}(y_i - \hat{y_i}, i=1,2,...,n)$, where MAD($\cdot$) is the median absolute deviation adjusted by a factor of 1.4826, and $\hat{y_i}$ is the estimated value of $y_i$. Compared with the conventional least-square loss function with $\rho(t)=t^2$, it can alleviate the influence of a single sample, especially the sample with a large noise, on estimating the residuals, and thus has less dependence on the distribution of noise and is more robust to the noise. 

The second term is a penalty, which is responsible for sparse estimation and selection of relevant genes within each subgroup. The $p$ genes have been organized into pre-specified $L$ overlapping clusters. The genes within a cluster share common biological functions and tend to behave coordinately. Thus, if a gene is relevant to the phenotype of interest, then genes in the same cluster may also be relevant. Massive researches have shown that, among tens of thousands of genes, only dozens of, or hundreds of, genes play leading roles in the occurrence and development of diseases \cite{Van2002}. The "sparsity" trait of genes renders that not all genes in a cluster are relevant, though these genes are similar to each other \cite{Gu2012Centrality}. We would like to identify the particularly important genes in clusters of interest. To this end, we adopt a sparse overlapping group lasso penalty (SOGlasso) \cite{Rao2016Classification}, which is defined as  
\begin{equation}
\label{eq:penalty}
	\phi(\beta_k; \gamma_k,\lambda_k)=\lambda_k\inf_{v_k^{(l)}\in V_l,\sum_{l=1}^L v_k^{(l)}=\beta_k}\sum_{l=1}^L \left[ \gamma_k\|v_k^{(l)}\|_1 + (1-\gamma_k)\omega_l \Vert v_k^{(l)}\Vert_2\right]. 
\end{equation}
Here $V_l=\{v\in\mathbb{R}^p|v_j=0 \ \text{if}\ j\notin C_l\}$; $\omega_l>0$ ($l=1,\ldots, L$) is the predefined weight for cluster $L$. Following Yuan and Lin \cite{yuan2006}, we set $\omega_l = \sqrt{p_l}$. 
The tuning parameter $\gamma_k\in [0,1]$ tradeoffs the contribution of $l_2$ and $l_1$ norm terms per gene cluster. The $l_2$ penalty promotes selecting only a few of the clusters, and the $l_1$ penalty promotes selecting a few of genes within a cluster. The tuning parameter $\lambda_k>0$ controls the overall strength of penalization. We allow the tuning parameters to be subgroup-specific so as to enhance the flexibility of penalization strength, thereby improving the accuracy in the identification of important genes. When the clusters in $\mathcal{C}$ do not overlap, the SOGlasso penalty reduces to the sparse group lasso \cite{Simon2013}. If $\gamma_k=0$, then the penalty (\ref{eq:penalty})  reduces to the overlapping group lasso \cite{Jacob2009}. Although the overlapping group lasso can select some important gene clusters, it may be too constraining because it forces all the genes in a cluster to be active at once: if a gene is selected, then all the genes within that cluster are selected. If the cluster structure of genes is ignored, namely each cluster $C_l$ is a singleton, then the penalty (\ref{eq:penalty}) reduces to the standard lasso. Though the lasso can select important individual gene, it completely ignores the underlying cluster structure of genes, and thus may be too under-constrained. As an interpolation between the overlapping group lasso and the lasso, the SOGlasso penalty realizes the mutual aided of the two types of penalties: it can select not only important clusters but also the particular important genes within these clusters. 

\subsection{Computation}
The optimization problem in Eq. (\ref{eq:obj}) involves minimizing a penalized objective function with respect to all possible groupings of $n$ samples. The number of partitions of $n$ samples into $K$ subgroups increases steeply with $n$, making exhaustive search virtually impossible. To overcome the hurdle, we adopt an iterative procedure by a similar idea as $K$-means clustering. The proposed algorithm is summarized in Algorithm \ref{alg1}.

\begin{algorithm}
\caption{Iterative strategy for solving Eq. (\ref{eq:obj})}\label{alg1}
\begin{algorithmic}
\State{\textbf{Input:}} dataset $\{X_i,y_i\}_{i=1}^n$, number of subgroups $K$, and number of starting partitions $R$. 
\State Initialize the lowest objective function value $l^*=10^5$, the optimal partition $\mathcal{G}^*=null$, and the optimal coefficients $\bm \beta^*=\bm 0$.  
\For {$r=1$ {\textbf{to}} $R$}
\State Set $t=0$. Randomly assign $n$ samples into $K$ disjoint subgroups and initialize the subgroup membership $\mathcal{G}(t)$. 
\Repeat 
\State 1: \textbf{[Update]} For $k \in \{1,...,K\}$, compute 
\begin{equation}
	\label{eq:algo}
\beta_k(t+1)=\mathop{\arg\min}_{\beta\in \mathbb{R}^p}\sum_{\{i|g_i(t)=k\}}\rho_\delta(y_i-X_i^{\top}\beta_k) + \phi(\beta_k;\gamma_k,\lambda_k).
\end{equation}
\State  2: \textbf{[Assignment]} For $i \in \{1,2,...,n\}$, compute 
\[g_i(t+1) = \mathop{\arg\min}_{k\in\{1,2,...,K\}}\rho_\delta(y_i-X_i\beta_k(t+1)),\]
\State where we take the minimum $k$ in case of a non-unique solution. 
\State 3: $t=t+1$ 
\Until {the objective function $\mathcal{L}(\bm\beta,\mathcal{G}) $ in Eq. (\ref{eq:obj}) converges.} 
\If {$\mathcal{L}(\bm\beta(t),\mathcal{G}(t))<l^*$} 
\State $l^*=\mathcal{L}(\bm\beta(t),\mathcal{G}(t))$, $\mathcal{G}^*=\mathcal{G}(t)$, $\bm \beta^*=\bm \beta(t)$
\EndIf
\EndFor
\State{\textbf{Output:}} $l^*$, $\mathcal{G}^*$, and $\bm\beta^*$.
\end{algorithmic}  
\end{algorithm}

The algorithm shares the similar spirit with the classic K-means clustering in the sense that it alternates between the "update" step and "assignment" step in each round of iteration. In the update step (Step 2), each $\beta_k$ is solved separately. In the numerical study, we, first, employ the covariate duplication method \cite{Jacob2009} to convert the problem to the non-overlapping sparse group lasso in an expanded space, and then use the block coordinate descent algorithm, which is a well-developed algorithm for tackling high-dimensional data, to estimate the subgroup-specific coefficients. The detailed calculation is provided in Appendix \ref{app:alg}. In the assignment step (Step 3), each individual sample $i$ is assigned to the subgroup $g_i$ with the shortest distance, where the distance is measured by the Huber loss function. The algorithm is terminated when the change of the objective function values between two successive iterations is less than $10^{-3}$. The objective function is non-increasing in the number of iterations, and convergence is typically very fast. In our numerical study, the algorithm can converge within 50 iterations in most repetitions. However, since the objective function in Eq. (\ref{eq:obj}) is non-convex, the solution is dependent on the starting values. To mitigate the dependence of the algorithm on the initial partition, we draw multiple starting partitions randomly and choose the one yielding the lowest objective function value. Our simulation results show that this practice works well. We also consider a more delicate initialization technique, namely variable neighborhood search (VNS) method \cite{Bonhomme2015Grouped}, which adopts local searching and neighborhood jumping ideas. Numerical experiments show that VNS leads to similar results, but with greater computational cost. Comprehensively considering effectiveness and performance cost, we choose to adopt the multiple initial values technique in the proposed approach.

\subsection{Implementation}  \label{sec:imp}
The proposed approach involves tuning parameters: regularization parameters $\lambda_k$'s, convex combination parameters $\gamma_k$'s as well as the number of subgroups $K$. For $\lambda_k$'s and $\gamma_k$'s, cross-validation (CV) over a grid search is the commonly adopted method to select the optimal combination of parameters, but this becomes increasingly prohibitive with the increase of $K$. Therefore, instead of first performing penalized linear regression for given $\lambda_k$ and $\gamma_k$ and then searching for the optimal combination of $\lambda_k$ and $\gamma_k$, we propose conducting tuning of $\lambda_k$ and $\gamma_k$ with CV inside the iterations of Algorithm \ref{alg1}. Specifically, under Algorithm \ref{alg1}, all the subgroups are disentangled at the update step, we could hence perform parameter tuning inside each update step within each subgroup. This is to say, at the update step, we not only estimate the regression coefficients but also find the best tuning parameters $\lambda_k$ and $\gamma_k$ for each subgroup. In practice, we adopt the efficient five-fold CV for selecting the optimal $\lambda_k$ and $\gamma_k$. More precisely, we partition each subgroup $k$ randomly into five non-overlapping subsets with equal sizes. We apply a two-dimensional grid search for $\lambda_k$ and $\gamma_k$ with $\gamma_k\in\{0.1,0.3,0.5,0.7\}$. For $\lambda_k$, we start with the smallest value $\lambda_k^{\max}$ under which all regression coefficients are zero, and then gradually reduces $\lambda_k$ to $\lambda_k^{\min}=10^{-3}\lambda_k^{\max}$ over a grid of values. The grid is 20 equi-distant on the log-scale. Since we no longer need to run the algorithm multiple times on a $K$-dimensional grid space of the penalty parameters, we could largely reduce the computational time. For the selection of the number of subgroups $K$, we use a modified BIC criterion \cite{wang2007} that minimizes 
\[
	\text{BIC}(K) = \text{log}\left[\sum_{k=1}^K\sum_{\{i|g_i=k\}}\rho_\delta(y_i-X_i^{\top}\beta_k)/n\right]+C\text{log}(n)/n \times d_f
\]
with respect to $K$, where $d_f$ is the total number of nonzero regression coefficients across $K$ subgroups, and $C=\log(\log(pK))$ \cite{Wang2009}. 

To implement the proposed approach, we have developed a Python code and made it publicly at https://github.com/YifanSun/ROG-Subgroup. The proposed approach is computationally affordable. For example, with a fixed number of subgroups $K$, for a simulation example with $n=300$, $p=200$, and two subgroups, the analysis can be accomplished within 40 seconds using a laptop with Intel(R) Core(TM) i5-7267U @ 3.10GHz CPU cores and 8G RAM.  

\section{Numerical simulation}
\label{sec:sim}
In this section, we evaluate the performance of the proposed approach through simulation study. We use different sample sizes by letting $n=300$ and $n=500$. The $n$ samples belong to two subgroups. We consider two subgroup structures: (a) balanced, where the two subgroups have the same sizes; and (b) unbalanced, where the two subgroups have sizes in a ratio of 7:3. To simulate gene expression data, a total of $p=200$ covariates are generated from a multivariate normal distribution with marginal means 0 and variances 1. We consider an auto-regression correlation structure, where the $j$th and $k$th genes have correlation $0.5^{|j-k|}$. With respect to the cluster structure of genes, we consider the following. 

\noindent {\bf Equal size} Each cluster entails ten  genes. For the overlapping pattern of clusters, we consider three scenarios: 

\noindent \underline{(S1)} Mild overlapping. These genes belong to 24 clusters with 2 genes of overlap between two successive clusters: $\{1,\ldots,10\}, \{9,\ldots,18\},\ldots, \{190,\ldots,199\}$;

\noindent\underline{(S2)} Severe overlapping. These genes belong to 39 clusters with 5 genes of overlap between two successive clusters: $\{1,\ldots, 10\}, \{5,\ldots, 15\},\ldots, \{191,\ldots, 200\}$;

\noindent \underline{(S3)} Moderate overlapping. These genes belong to 24 clusters with a varied number of overlapped genes between two successive clusters. Specifically, 12, 6 and 8 clusters have 2, 5, and 0 overlapped genes, respectively, with their preceding clusters. 
\[
	\{1,\ldots,10\}, \{9,\ldots,18\},\ldots, \{97,\ldots,106\} 
	\llap{$\overbrace{\phantom{\{1,\ldots,10\}, \{9,\ldots,18\},\ldots, \{97,\ldots,106\}}}^{\text{12 clusters with 2 overlaps}}$}
	 ,\{102,\ldots,111\},\ldots,\{122,\ldots,131\}
	\llap{$\underbrace{\phantom{ \{97,\ldots,106\}, \{102,\ldots,111\},\ldots,\{122,\ldots,131\}}}_{\text{6 clusters with 5 overlaps}}$}
	 ,\{132,\ldots,141\},\ldots, \{192,\ldots,201\}
	\llap{$\overbrace{\phantom{ \{122,\ldots,131\},\{132,\ldots,141\},\ldots, \{192,\ldots,201\}}}^{\text{8 clusters with 0 overlaps}}$}.
\]
\noindent{\bf Unequal size} The sizes of the clusters range from 1 to 20. For the overlapping pattern of clusters, we consider three scenarios: 

\noindent \underline{(S4)} Mild overlapping. These genes belong to 36 clusters, of which 8 clusters have only one gene, 7 clusters have 3 genes, 7 clusters have 5 genes, 7 clusters have 10 genes, 4 clusters have 15 genes, and 3 clusters have 20 genes. There are two overlapped genes between two successive clusters that have more than two genes: $\{1,\ldots,5\}, \{4,\ldots,8\}, \{7,\ldots,16\}, \{15,\ldots,24\}, \{23, 24, 25\}, \{24, 25, 26\}, \{25,\ldots,39\}, \{38,\ldots,52\}, \{51,\ldots,70\}, \{69, 70, 71\},\\
,\ldots,\{73, 74, 75\}, \{74,\ldots,78\}, \ldots, \{86,\ldots,90\}, \{89,\ldots,98\},\ldots, \{121,\ldots,130\}, \{129,\ldots,143\},\ldots, \{173,\ldots,192\}, \{193\},\\\ldots, \{200\}$.

\noindent \underline{(S5)} Severe overlapping. These genes belong to 38 clusters, of which 10 clusters have only one gene, 5 clusters have 6 genes, 5 clusters have 8 genes, 7 clusters have 10 genes, 7 clusters have 15 genes, and 4 clusters have 20 genes. There are 5 overlapped genes between two successive clusters that have more than 5 genes: $\{1,\ldots,6\}, \{2,\ldots,7\}, \{3,\ldots,10\}, \{6,\ldots,13\}, \{9,\ldots,18\}, \{14,\ldots,23\}, \{19,\ldots,33\}, \{29,\ldots,43\}, \{39,\ldots,58\}, \{54,\ldots,73\},\ldots, \\
\{71,\ldots,76\}, \{72,\ldots,79\},\ldots, \{78,\ldots,85\}, \{81,\ldots,90\},\ldots, \{91,\ldots,100\},\ldots, \{101,\ldots,110\}, \{106,\ldots,120\},\ldots, \{146,\ldots,\\
160\}, \{156,\ldots,175\},\{171,\ldots,190\}, \{191\},\ldots,\{200\}$.

\noindent \underline{(S6)} Moderate overlapping. These genes belong to 35 clusters, of which 10 clusters have only one gene, 2 clusters have 3 genes, 5 clusters have 5 genes, 3 clusters have 6 genes, 3 clusters have 8 genes, 4 clusters have 10 genes, 4 clusters have 15 genes, and 4 clusters have 20 genes. Among them, 18, 8 and 11 clusters have 2, 5, and 0 overlapped genes, respectively, with their respective preceding clusters: $\{1, 2, 3\}, \{2, 3, 4\}, \{3,\ldots,7\}, \{6,\ldots,10\}, \{9,\ldots,18\}, \{17,\ldots,26\}, \{25,\ldots,39\}, \{38,\ldots, 52\}, \{51,\ldots,70\}, \{69,\ldots,88\}, \{87,\ldots,\\
91\}, \ldots, \{93,\ldots,97\}, \{96,\ldots,101\},\ldots, \{104,\ldots,109\}, \{108,\ldots,115\},\ldots,\{120,\ldots,127\}, \{126,\ldots,135\}, \{131,\ldots,140\}, \\\{136,\ldots,150\}, \{146,\ldots,160\}, \{156,\ldots, 175\}, \{171,\ldots,190\}, \{191\},\ldots, \{200\}$. 

Each subgroup has 15 important genes. Specifically, in the first subgroup, the set of important genes is $\{1, 3, 6, 9, 10, 11, 13, 16, 20, 21, 25, 27, 29, 31, 33\}$, and their coefficients are $\{2, 1, 0.5, -1, 1.5, 0.5, -1, 2, -1, 0.5, -1, 0.5, 1.5, 0.5, 1\}$; in the second subgroup, the set of important genes is \{2, 3, 5, 9, 10, 12, 14, 17, 18, 19, 22, 23, 26, 31, 33\}, and their coefficients are \{-2, 1, -2, -1, 0.5, 1.5, 1, -1, -0.5, 2, -0.5, 1, -2, -0.5, -1\}. The two subgroups share five important genes: \{3, 9, 10, 31, 33\}, of which the 3rd, 9th, and 31st genes have the identical coefficients in two subgroups. The response variables are generated from the heterogeneous regression model $y_i=X_i\beta_{g_i}+0.5\varepsilon_i$ for $i=1,\ldots,n$, where the error terms $\varepsilon_i$'s have the following distributions: (1) $t(1)$, (b) $0.7N(0, 1) + 0.3t(1)$, and (c) $N(0,1)$. 

Besides the proposed robust overlapping clusters approach (denoted as R-OC), we also consider the following alternatives for comparison. 
The robust unstructured approach (denoted as R-US) has the same modeling framework as the proposed; however, the cluster structure of genes is ignored. The penalized loss function of R-US is $\sum_{k=1}^K\sum_{\{i|g_i=k\}}\rho_{\delta}(y_i-X_i^{\top}\beta_k)+\sum_{k=1}^K\lambda_k\Vert\beta_k\Vert_1$. We also consider the nonrobust alternatives NR-OC and NR-US, where the Huber lasso $\rho_{\delta}(\cdot)$ in the penalized loss function of the robust approaches is replaced by the nonrobust least-square loss. The tuning parameters of these alternatives are determined by five-fold CV inside iterations, in the same way as in the proposed R-OC. For each approach, we use 20 different starting values and choose the one giving the smallest objective function to mitigate the dependency of the algorithm on initial values. We acknowledge that there are other heterogeneity analysis approaches that can be adopted to analyze the simulated data. The above three alternatives are the most direct competitors as they follow the similar framework as the proposed approach. In particular, the robustness of the proposed approach can be demonstrated by comparing with NR-OC and NR-US as well. By comparing with R-US, the advantages of accounting for the overlapping cluster structure of genes of the proposed approach can be directly established. We have briefly experimented with penalized FMRs \cite{Khalili2007Variable, Lloyd2018} but found unacceptable results, hence such methods are not reported.

In evaluation, we are interested in the identification of subgroups and important genes. To assess the subgroup identification performance, we consider three measures, namely the estimated number of subgroups ($\hat{K}$), the percentage of $\hat{K}$ equaling the true number of subgroups, and the Adjusted Rand Index (ARI), where ARI measures the agreement between the structure of the estimated subgroups and that of the true subgroups. We evaluate the gene identification performance by using the True Positive Rate (TPR), the False Positive Rate (FPR), and the Matthews Correlation Coefficient (MCC)\cite{Matthews1975}. The MCC is comprehensive evaluation index to measure the quality of identifying nonzero effects, which has been widely used in the field of statistics and bioinformatics. The MCC has a range of -1 to 1, where -1 indicates a completely wrong identification while 1 indicates a completely correct identification.  In addition, we also evaluate estimation performance. Specifically, estimation performance is measured by the Root Mean Squared Errors (RMSE), which is defined as $\sqrt{\sum_{i=1}^n\Vert\hat{\beta}_{\hat{g}_i}-\beta_{g_i}\Vert_2^2/n}$. 

Summary statistics are computed based on 100 independent replicates. Table \ref{tab:kselect} reports the sample mean, median, and standard error of the estimated number of subgroups $\hat{K}$ as well as the percentage that $\hat{K}$ equals to the true number of subgroups by the proposed R-OC under the balanced subgroup structure. The median of $\hat{K}$ is 2 which is the true number of subgroups for all settings. As $n$ increases, the mean reaches closer to 2, the standard error becomes smaller, and the percentage of correctly determining the number of subgroups becomes larger. We also examine the unbalanced subgroup structure setting and observe similar patterns. 
In the following, we set the number of subgroups as the true $K$ for all approaches. 

\begin{table}[]\small
\centering
\caption{Simulation results: the sample mean, median, standard error (s.e.) of $\hat{K}$ and the percentage of $\hat{K}$ equaling the true number of subgroups ($K=2$) by the proposed R-OC for balanced subgroup structure.}
\label{tab:kselect}
\resizebox{\textwidth}{!}{%
\begin{tabular}{ccccccccccc}
\hline
\multirow{2}{*}{}   & \multirow{2}{*}{Error} & \multicolumn{4}{c}{$n=300$}         &  & \multicolumn{4}{c}{$n=500$}         \\ \cline{3-6} \cline{8-11} 
                    &                        & Mean & Median & s.e. & Percentage &  & Mean & Median & s.e. & Percentage \\ \hline
\multirow{3}{*}{S1} & $t(1)$                   & 1.67 & 2.00   & 0.51 & 0.75       &  & 1.96 & 2.00   & 0.33 & 0.96       \\
                    & $0.7N(0,1)+0.3t(1)$      & 2.05 & 2.00   & 0.32 & 0.93       &  & 2.02 & 2.00   & 0.23 & 0.97       \\
                    & $N(0,1)$                 & 2.26 & 2.00   & 0.54 & 0.71       &  & 2.08 & 2.00   & 0.34 & 0.92       \\ 
                    \\
\multirow{3}{*}{S2} & $t(1)$                   & 1.68 & 2.00   & 0.5  & 0.76       &  & 1.97 & 2.00   & 0.33 & 0.97       \\
                    & $0.7N(0,1)+0.3t(1)$      & 2.07 & 2.00   & 0.33 & 0.92       &  & 2.03 & 2.00   & 0.22 & 0.97       \\
                    & $N(0,1)$                 & 2.29 & 2.00   & 0.55 & 0.70       &  & 2.10 & 2.00   & 0.35 & 0.90       \\ 
                    \\
\multirow{3}{*}{S3} & $t(1)$                   & 1.70 & 2.00   & 0.49 & 0.78       &  & 1.97 & 2.00   & 0.32 & 0.96       \\
                    & $0.7N(0,1)+0.3t(1)$      & 2.04 & 2.00   & 0.33 & 0.93       &  & 2.02 & 2.00   & 0.22 & 0.97       \\
                    & $N(0,1)$                & 2.27 & 2.00   & 0.55 & 0.71       &  & 2.09 & 2.00   & 0.35 & 0.91       \\
                    \\
\multirow{3}{*}{S4} & $t(1)$                   & 1.68 & 2.00   & 0.50 & 0.76       &  & 1.97 & 2.00   & 0.33 & 0.97       \\
                    & $0.7N(0,1)+0.3t(1)$      & 2.06 & 2.00   & 0.33 & 0.92       &  & 2.03 & 2.00   & 0.22 & 0.97       \\
                    & $N(0,1)$                 & 2.28 & 2.00   & 0.55 & 0.70       &  & 2.09 & 2.00   & 0.35 & 0.91       \\ 
                    \\
\multirow{3}{*}{S5} & $ t(1) $                   & 1.69 & 2.00   & 0.49 & 0.77       &  & 1.97 & 2.00   & 0.32 & 0.96       \\
                    & $0.7N(0,1)+0.3t(1)$      & 2.05 & 2.00   & 0.33 & 0.93       &  & 2.02 & 2.00   & 0.22 & 0.97       \\
                    & $N(0,1)$                 & 2.28 & 2.00   & 0.55 & 0.71       &  & 2.09 & 2.00   & 0.35 & 0.91       \\ 
                    \\ 
\multirow{3}{*}{S6} & $t(1)$                   & 1.69 & 2.00   & 0.50 & 0.77       &  & 1.97 & 2.00   & 0.33 & 0.96       \\
                    & $0.7N(0,1)+0.3t(1)$      & 2.05 & 2.00   & 0.33 & 0.93       &  & 2.02 & 2.00   & 0.22 & 0.97       \\
                    & $N(0,1)$                 & 2.27 & 2.00   & 0.55 & 0.71       &  & 2.09 & 2.00   & 0.35 & 0.91       \\ \hline
\end{tabular}%
}
\end{table}

For the settings with $n=300$ and balanced subgroup structure, we show the subgrouping, gene identification, and estimation performance in Table \ref{tab:n3bal1} for scenarios S1-S3 and in Table \ref{tab:n3bal2} for scenarios S4-S6. In general, the proposed R-OC performs the best in subgrouping in the majority of the scenarios. When there is no contamination (error $N(0,1)$), R-OC and its nonrobust counterpart NR-OC perform comparably well. It is noted that, in some scenarios, for example, scenario S2 (Table \ref{tab:n3bal1}), R-OC performs slightly better than NR-OC. A possible reason is that the robust Huber loss function can alleviate the influence of a misclassified sample, which usually produces a large error, in estimating the residuals, and thereby it may improve, to some degree, the estimation of subgroup-specific coefficients and subgrouping accuracy as well.  When data have contamination, R-OC significantly outperforms the alternatives. For example, in scenario S1, under error $0.7N(0,1)+0.3t(1)$ (Table \ref{tab:n3bal1}), R-OC has ARI=0.695, compared to 0.106 (NR-US), 0.129 (NR-OC), and 0.338 (R-US). Under the high-dimensional situation, the subgrouping and gene identification are not isolated, but dependent upon each other. As a result, though R-US also adopts the robust Huber loss function, but because it ignores the cluster structure of genes, which impairs the accuracy of gene identification, it performs poorly in subgrouping. NR-US has the worst performance among all the approaches. This result is as expected since NR-US neither adopts robust loss function nor taking the cluster structure of genes into account. We also conduct additional simulations to examine the dependence of the proposed approach on the number of starting values. Figure \ref{fig:init} in Appendix \ref{sec:appB} shows the ARI as a function of the number of starting values $\kappa$ under scenario S1, balanced structure, $n=300$. It can be observed that all approaches perform better with larger $\kappa$, and the proposed R-OC outperforms the alternatives regardless of $\kappa$, indicating that the merit of R-OC in subgrouping is insensitive to the number of starting values.  

For gene identification, the proposed R-OC again has the best or close to the best performance compared to the alternatives. 
NR-OC achieves similar performance with R-OC when there is no contamination, but it clearly suffers when data have contamination. For example, in scenario S5, under error $N(0,1)$ (Table \ref{tab:n3bal1}), NR-OC has (TPR, FPR, MCC)=(1.000, 0.007, 0.966) approaching that of R-OC: (TPR, FPR, MCC)=(1.000, 0.005, 0.966); however, when the error distribution is $t(1)$, NR-OC has (TPR, FPR, MCC)=(0.067, 0.015, -0.009), which is significantly inferior to R-OC: (TPR, FPR, MCC)=(0.802, 0.044, 0.635). Comparison with the robust unstructured approach R-US also shows the advantage of the proposed approach. R-US identifies fewer true positives and more false positives, leading to a lower MCC. Across all scenarios, R-US and NR-US perform worse than R-OC, and as in subgrouping, NR-US has the worst performance. In addition, it is observed that, without contamination, the proposed R-OC has estimation performance comparable to NR-OC and outperforms the robust alternative R-US. With contamination, R-OC has significantly smaller prediction errors. For example, in scenario S6, under error $0.7N(0,1)+0.3t(1)$ (Table \ref{tab:n3bal2}), R-OC has RMSE=0.146, compared to 0.493 (NR-US), 0.410 (NR-OC), and 0.381 (R-US). The results under other settings are reported in Tables \ref{tab:n5bal1}-\ref{tab:n5unbal2} (Appendix \ref{sec:appB}). Similar patterns have been observed. 

We also consider three subgroups case. Specifically, suppose $n=300$ samples form three subgroups of equal size. The set of important genes and their coefficients in the first two subgroups are the same as that in the aforementioned two subgroups case. The third subgroup also has 15 important genes: \{0, 3, 4, 6, 7, 10, 11, 12, 15, 17, 24, 25, 31, 32, 34\}, of which the coefficients are \{-1, 1, -2, -0.5, 1, 2, -1, 2, 0.5, -1, 1, 1.5, 0.5, 1, -1.5\}. The three subgroups share three important genes: \{3,10,31\}, where the 3rd gene has the identical coefficient in three subgroups. The other settings are the same as describe above. We use scenario S1 as a representative and present results in Tables \ref{tab:k3select}-\ref{tab:k3} (Appendix \ref{sec:appB}). The proposed R-OC is again observed to have favorable performance in subgrouping, gene identification, and coefficient estimation as well. 

Finally, we apply the proposed R-OC to the low-dimensional data. We consider the balanced subgroup structure with $K=2$, $n=300$ and $p=10,20,32,50$. The overlapping cluster structure of genes and their coefficients are provided in Appendix \ref{sec:appB-}. 
For the low-dimensional data, we add the fusion method \cite{Ma2020Exploration} as an alternative. Tables \ref{tab:penalized_fusion_n}-\ref{tab:penalized_fusion_t} (Appendix \ref{sec:appB}) present the simulation results. It can be seen that the proposed R-OC outperforms the fusion method, and still has the best or close to the best performance.

\begin{table}[]\small
\centering
\caption{Simulation scenarios S1-S3 for balanced subgroup structure with $n=300$. Each cell shows the median (median absolute deviation). }
\label{tab:n3bal1}
\resizebox{\textwidth}{!}{%
\begin{tabular}{ccccccc}
\hline
Error                              & Method & ARI                   & TPR                   & FPR                   & MCC                   & RMSE                  \\ \hline
\multicolumn{7}{c}{S1}                                                                                                                                              \\
\multirow{4}{*}{$t(1)$}              & NR-US  & 0.002(0.011)          & 0.067(0.109)          & \textbf{0.008(0.084)} & 0.095(0.071)          & 0.410(0.132)          \\
                                   & NR-OC  & 0.005(0.007)          & 0.083(0.154)          & 0.035(0.130)          & 0.051(0.112)          & 0.403(0.118)          \\
                                   & R-US   & 0.090(0.088)          & 0.383(0.102)          & 0.138(0.085)          & 0.159(0.137)          & 0.367(0.046)          \\
                                   & R-OC   & \textbf{0.484(0.124)} & \textbf{0.933(0.170)} & 0.024(0.019)          & \textbf{0.788(0.158)} & \textbf{0.198(0.044)} \\
                                   &        & \textbf{}             & \textbf{}             &                       & \textbf{}             & \textbf{}             \\
\multirow{4}{*}{$0.7N(0,1)+0.3t(1)$} & NR-US  & 0.106(0.134)          & 0.300(0.154)          & 0.029(0.105)          & 0.306(0.077)          & 0.325(0.151)          \\
                                   & NR-OC  & 0.129(0.109)          & 0.467(0.291)          & 0.032(0.197)          & 0.412(0.121)          & 0.306(0.167)          \\
                                   & R-US   & 0.338(0.191)          & 0.733(0.282)          & 0.056(0.078)          & 0.649(0.223)          & 0.226(0.138)          \\
                                   & R-OC   & \textbf{0.695(0.038)} & \textbf{1.000(0.102)} & \textbf{0.011(0.008)} & \textbf{0.900(0.174)} & \textbf{0.154(0.050)} \\
                                   &        & \textbf{}             & \textbf{}             & \textbf{}             & \textbf{}             & \textbf{}             \\
\multirow{4}{*}{$N(0,1)$}            & NR-US  & 0.789(0.338)          & \textbf{1.000(0.131)} & 0.019(0.169)          & 0.892(0.307)          & 0.116(0.140)          \\
                                   & NR-OC  & \textbf{0.809(0.210)} & \textbf{1.000(0.088)} & \textbf{0.003(0.091)} & \textbf{0.982(0.202)} & \textbf{0.102(0.070)} \\
                                   & R-US   & 0.795(0.251)          & \textbf{1.000(0.101)} & 0.012(0.189)          & 0.928(0.207)          & 0.125(0.140)          \\
                                   & R-OC   & \textbf{0.809(0.030)} & \textbf{1.000(0.003)} & 0.005(0.003)          & 0.966(0.021)          & \textbf{0.102(0.010)} \\ \hline
\multicolumn{7}{c}{S2}                                                                                                                                              \\
\multirow{4}{*}{$t(1)$}              & NR-US  & -0.001(0.009)         & 0.050(0.098)          & 0.007(0.075)          & 0.095(0.056)          & 0.402(0.092)          \\
                                   & NR-OC  & 0.001(0.005)          & 0.067(0.102)          & 0.014(0.078)          & 0.099(0.084)          & 0.480(0.186)          \\
                                   & R-US   & 0.027(0.053)          & 0.300(0.173)          & 0.130(0.076)          & 0.139(0.098)          & 0.368(0.031)          \\
                                   & R-OC   & \textbf{0.438(0.218)} & \textbf{0.867(0.177)} & 0.030(0.040)          & \textbf{0.750(0.253)} & \textbf{0.205(0.080)} \\
                                   &        & \textbf{}             & \textbf{}             &                       & \textbf{}             & \textbf{}             \\
\multirow{4}{*}{$0.7N(0,1)+0.3t(1)$} & NR-US  & 0.098(0.123)          & 0.270(0.173)          & 0.087(0.117)          & 0.179(0.094)          & 0.360(0.185)          \\
                                   & NR-OC  & 0.103(0.139)          & 0.333(0.270)          & 0.033(0.162)          & 0.343(0.108)          & 0.322(0.151)          \\
                                   & R-US   & 0.322(0.175)          & 0.653(0.213)          & 0.095(0.092)          & 0.524(0.213)          & 0.230(0.148)          \\
                                   & R-OC   & \textbf{0.606(0.068)} & \textbf{1.000(0.123)} & \textbf{0.011(0.020)} & \textbf{0.886(0.202)} & \textbf{0.167(0.056)} \\
                                   &        & \textbf{}             & \textbf{}             & \textbf{}             & \textbf{}             & \textbf{}             \\
\multirow{4}{*}{$N(0,1)$}            & NR-US  & 0.750(0.355)          & \textbf{1.000(0.136)} & 0.008(0.121)          & 0.950(0.371)          & 0.119(0.119)          \\
                                   & NR-OC  & 0.774(0.254)          & \textbf{1.000(0.088)} & 0.008(0.135)          & 0.950(0.264)          & 0.113(0.086)          \\
                                   & R-US   & \textbf{0.785(0.166)} & \textbf{1.000(0.103)} & 0.008(0.071)          & 0.950(0.182)          & 0.112(0.059)          \\
                                   & R-OC   & \textbf{0.785(0.039)} & \textbf{1.000(0.014)} & \textbf{0.005(0.005)} & \textbf{0.966(0.027)} & \textbf{0.110(0.011)} \\ \hline
\multicolumn{7}{c}{S3}                                                                                                                                              \\
\multirow{4}{*}{$t(1)$}              & NR-US  & 0.001(0.01)           & 0.033(0.090)          & \textbf{0.007(0.052)} & 0.063(0.058)          & 0.416(0.129)          \\
                                   & NR-OC  & 0.002(0.01)           & 0.060(0.102)          & 0.023(0.109)          & 0.064(0.048)          & 0.479(0.108)          \\
                                   & R-US   & 0.015(0.075)          & 0.417(0.176)          & 0.163(0.074)          & 0.163(0.098)          & 0.380(0.038)          \\
                                   & R-OC   & \textbf{0.391(0.159)} & \textbf{0.883(0.269)} & 0.033(0.039)          & \textbf{0.746(0.240)} & \textbf{0.215(0.057)} \\
                                   &        & \textbf{}             & \textbf{}             &                       & \textbf{}             & \textbf{}             \\
\multirow{4}{*}{$0.7N(0,1)+0.3t(1)$} & NR-US  & 0.010(0.076)          & 0.250(0.170)          & 0.125(0.107)          & 0.108(0.111)          & 0.422(0.072)          \\
                                   & NR-OC  & 0.012(0.155)          & 0.290(0.233)          & 0.137(0.188)          & 0.117(0.198)          & 0.426(0.147)          \\
                                   & R-US   & 0.231(0.243)          & 0.567(0.221)          & 0.132(0.114)          & 0.310(0.267)          & 0.301(0.094)          \\
                                   & R-OC   & \textbf{0.666(0.100)} & \textbf{1.000(0.095)} & \textbf{0.008(0.006)} & \textbf{0.942(0.097)} & \textbf{0.142(0.028)} \\
                                   &        & \textbf{}             & \textbf{}             & \textbf{}             & \textbf{}             & \textbf{}             \\
\multirow{4}{*}{$N(0,1)$}            & NR-US  & 0.756(0.313)          & \textbf{1.000(0.202)} & 0.008(0.133)          & 0.950(0.338)          & 0.119(0.111)          \\
                                   & NR-OC  & \textbf{0.791(0.251)} & \textbf{1.000(0.131)} & \textbf{0.005(0.125)} & \textbf{0.966(0.274)} & \textbf{0.108(0.084)} \\
                                   & R-US   & 0.772(0.249)          & \textbf{1.000(0.168)} & 0.007(0.179)          & 0.950(0.384)          & 0.114(0.126)          \\
                                   & R-OC   & 0.785(0.046)          & \textbf{1.000(0.004)} & \textbf{0.005(0.003)} & \textbf{0.966(0.018)} & 0.111(0.013)          \\ \hline
\end{tabular}%
}
\end{table}

\begin{table}[]\small
\centering
\caption{Simulation scenarios S4-S6 for balanced subgroup structure with $n=300$. Each cell shows the median (median absolute deviation). }
\label{tab:n3bal2}
\resizebox{\textwidth}{!}{%
\begin{tabular}{ccccccc}
\hline
Error                                & Method & ARI                   & TPR                   & FPR                   & MCC                   & RMSE                  \\ \hline
\multicolumn{7}{c}{S4}                                                                                                                                                \\
\multirow{4}{*}{$t(1)$}              & NR-US  & -0.001(0.009)         & 0.033(0.122)          & 0.005(0.079)          & 0.039(0.070)          & 0.425(0.100)          \\
                                     & NR-OC  & 0.000(0.016)          & 0.067(0.111)          & 0.022(0.069)          & 0.003(0.073)          & 0.429(0.137)          \\
                                     & R-US   & 0.063(0.082)          & 0.411(0.165)          & 0.133(0.072)          & 0.187(0.118)          & 0.382(0.046)          \\
                                     & R-OC   & \textbf{0.438(0.195)} & \textbf{0.844(0.101)} & \textbf{0.032(0.084)} & \textbf{0.765(0.132)} & \textbf{0.167(0.036)} \\
                                     &        & \textbf{}             & \textbf{}             & \textbf{}             & \textbf{}             & \textbf{}             \\
\multirow{4}{*}{$0.7N(0,1)+0.3t(1)$} & NR-US  & 0.101(0.039)          & 0.167(0.168)          & 0.031(0.101)          & 0.176(0.089)          & 0.434(0.291)          \\
                                     & NR-OC  & 0.104(0.017)          & 0.133(0.170)          & 0.038(0.099)          & 0.137(0.068)          & 0.452(0.195)          \\
                                     & R-US   & 0.321(0.144)          & 0.767(0.149)          & 0.087(0.085)          & 0.515(0.219)          & 0.309(0.177)          \\
                                     & R-OC   & \textbf{0.632(0.108)} & \textbf{0.883(0.077)} & \textbf{0.015(0.089)} & \textbf{0.846(0.235)} & \textbf{0.151(0.070)} \\
                                     &        & \textbf{}             & \textbf{}             & \textbf{}             & \textbf{}             & \textbf{}             \\
\multirow{4}{*}{$N(0,1)$}            & NR-US  & 0.722(0.380)          & \textbf{1.000(0.255)} & 0.014(0.122)          & 0.921(0.399)          & 0.127(0.126)          \\
                                     & NR-OC  & 0.745(0.352)          & \textbf{1.000(0.202)} & \textbf{0.007(0.151)} & \textbf{0.958(0.385)} & 0.121(0.126)          \\
                                     & R-US   & 0.716(0.373)          & \textbf{1.000(0.191)} & 0.010(0.171)          & 0.942(0.388)          & 0.128(0.128)          \\
                                     & R-OC   & \textbf{0.762(0.226)} & \textbf{1.000(0.135)} & \textbf{0.007(0.140)} & \textbf{0.958(0.327)} & \textbf{0.116(0.109)} \\ \hline
\multicolumn{7}{c}{S5}                                                                                                                                                \\
\multirow{4}{*}{$t(1)$}              & NR-US  & -0.001(0.006)         & 0.033(0.106)          & 0.007(0.069)          & 0.061(0.058)          & 0.392(0.142)          \\
                                     & NR-OC  & -0.002(0.003)         & 0.067(0.107)          & 0.015(0.094)          & -0.009(0.044)         & 0.466(0.110)          \\
                                     & R-US   & 0.049(0.109)          & 0.400(0.184)          & 0.144(0.059)          & 0.157(0.134)          & 0.353(0.052)          \\
                                     & R-OC   & \textbf{0.403(0.175)} & \textbf{0.802(0.169)} & \textbf{0.044(0.071)} & \textbf{0.635(0.160)} & \textbf{0.189(0.045)} \\
                                     &        & \textbf{}             & \textbf{}             & \textbf{}             & \textbf{}             & \textbf{}             \\
\multirow{4}{*}{$0.7N(0,1)+0.3t(1)$} & NR-US  & 0.073(0.104)          & 0.300(0.197)          & 0.152(0.110)          & 0.105(0.117)          & 0.439(0.198)          \\
                                     & NR-OC  & 0.081(0.118)          & 0.302(0.179)          & 0.133(0.110)          & 0.128(0.101)          & 0.458(0.121)          \\
                                     & R-US   & 0.303(0.109)          & 0.510(0.151)          & 0.100(0.082)          & 0.312(0.128)          & 0.315(0.104)          \\
                                     & R-OC   & \textbf{0.507(0.217)} & \textbf{0.841(0.198)} & \textbf{0.042(0.107)} & \textbf{0.692(0.138)} & \textbf{0.152(0.116)} \\
                                     &        & \textbf{}             & \textbf{}             & \textbf{}             & \textbf{}             & \textbf{}             \\
\multirow{4}{*}{$N(0,1)$}            & NR-US  & 0.721(0.322)          & \textbf{1.000(0.203)} & 0.008(0.121)          & 0.956(0.332)          & 0.132(0.144)          \\
                                     & NR-OC  & \textbf{0.735(0.328)} & \textbf{1.000(0.171)} & 0.007(0.134)          & \textbf{0.966(0.328)} & \textbf{0.122(0.137)} \\
                                     & R-US   & 0.727(0.325)          & \textbf{1.000(0.197)} & 0.008(0.166)          & 0.958(0.325)          & 0.129(0.134)          \\
                                     & R-OC   & 0.731(0.208)          & \textbf{1.000(0.162)} & \textbf{0.005(0.143)} & \textbf{0.966(0.327)} & \textbf{0.122(0.127)} \\ \hline
\multicolumn{7}{c}{S6}                                                                                                                                                \\
\multirow{4}{*}{$t(1)$}              & NR-US  & 0.000(0.008)          & 0.030(0.114)          & \textbf{0.006(0.074)} & 0.076(0.064)          & 0.396(0.121)          \\
                                     & NR-OC  & 0.002(0.010)          & 0.062(0.109)          & 0.016(0.082)          & 0.081(0.059)          & 0.465(0.124)          \\
                                     & R-US   & 0.050(0.096)          & 0.364(0.175)          & 0.125(0.066)          & 0.153(0.126)          & 0.329(0.049)          \\
                                     & R-OC   & \textbf{0.377(0.185)} & \textbf{0.754(0.135)} & 0.035(0.078)          & \textbf{0.654(0.146)} & \textbf{0.161(0.041)} \\
                                     &        & \textbf{}             & \textbf{}             &                       & \textbf{}             & \textbf{}             \\
\multirow{4}{*}{$0.7N(0,1)+0.3t(1)$} & NR-US  & 0.077(0.072)          & 0.217(0.183)          & 0.088(0.106)          & 0.123(0.103)          & 0.493(0.245)          \\
                                     & NR-OC  & 0.082(0.068)          & 0.204(0.175)          & 0.082(0.105)          & 0.119(0.085)          & 0.410(0.158)          \\
                                     & R-US   & 0.280(0.127)          & 0.562(0.150)          & 0.085(0.084)          & 0.362(0.174)          & 0.381(0.141)          \\
                                     & R-OC   & \textbf{0.506(0.113)} & \textbf{0.864(0.188)} & \textbf{0.052(0.098)} & \textbf{0.674(0.187)} & \textbf{0.146(0.093)} \\
                                     &        & \textbf{}             & \textbf{}             & \textbf{}             & \textbf{}             & \textbf{}             \\
\multirow{4}{*}{$N(0,1)$}            & NR-US  & 0.694(0.351)          & 0.912(0.229)          & 0.010(0.122)          & 0.896(0.366)          & 0.137(0.135)          \\
                                     & NR-OC  & 0.726(0.340)          & 0.922(0.187)          & 0.006(0.143)          & 0.917(0.357)          & 0.129(0.132)          \\
                                     & R-US   & 0.705(0.349)          & \textbf{0.932(0.194)} & 0.008(0.169)          & 0.912(0.357)          & 0.126(0.131)          \\
                                     & R-OC   & \textbf{0.730(0.217)} & \textbf{0.932(0.149)} & \textbf{0.005(0.142)} & \textbf{0.929(0.327)} & \textbf{0.117(0.118)} \\ \hline
\end{tabular}%
}
\end{table}

\section{Real data analysis}\label{sec:ccle}
The Cancer Cell Line Encyclopedia (CCLE) dataset contains 947 cancer cell lines from nine types of cancers with associated gene expression measurements and sensitivity score to 24 anti-cancer drugs \cite{Barretina2012}. We treat the sensitive score, that is, the area above the dose-response curve, as the response and use the gene expression measurements as covariates. Since these different types of cells are from distinct experimental and genetic backgrounds, it is expected that they may adopt different molecular mechanisms to cope with the attack of drugs. As such, our goal is to study the possible cell heterogeneity in their response to drug treatments, namely, to unveil the underlying subgroups of cells with respect to gene-sensitivity associations and identify the most important genes for each subgroup. Several heterogeneity analysis studies have been conducted on this data \cite{Dondelinger2020The, Chang2021Supervised}, although it is noted that they have adopted approaches significantly different from the proposed. 

After removing the cell lines with missing values, 400-500 samples are included for each drug. In the original data, there are 18, 899 gene expressions. We group these genes into (overlapping) pathways. Specifically, from the KEGG pathway database downloaded from Broad Institute, we identify 5245 unique genes, representing 186 pathways. We restrict the analysis to the 4431 genes which are in at least one pathway. Although, in principle, it is possible to directly apply the proposed approach on the 4431 genes, the small sample size $(400-500)$ and additional complexity brought by heterogeneity may lead to unstable and unreliable identifications and estimations \cite{Fan2008}. Thus, to obtain more reliable analysis results, we further conduct a marginal screening to screen out irrelevant genes and include 300 genes whose marginal associations with the response are the most significant. The final analyzed data contains 300 genes from 143 pathways. We treat each pathway as a gene cluster, where each gene cluster consists of the set of genes in a pathway. As shown in Figure \ref{fig:group_distr} (Appendix \ref{sec:appB}), these gene clusters entail a varied number of genes. The statistics of the 143 gene clusters are summarized as follows: the average number of genes in each cluster is 5.68, the largest gene cluster has  40 genes, and 98 genes appear in at least two clusters. 

We apply the proposed R-OC to each of the 24 drugs. For completeness, an intercept term is included in the model (\ref{eq:model}), which is also allowed to be subgroup-specific.
Take the drug Paclitaxel, an antimicrotubule drug to treat breast cancer, ovarian cancer, and lung cancer, as an example. The number of subgroups is determined by minimizing the BIC, in the same way as in simulation. The result is briefly presented in Figure \ref{fig: bic} (Appendix \ref{sec:appB}). Four subgroups, with sizes 67, 53, 147, and 67, are identified. Each subgroup entails cells from multiple types of cancers. We test the differences between these subgroups with respect to the proportions of different cancer types using Kruskal-Wallis test, and the resulted $p$-value is significant (<0.05). To investigate whether the subgroups are biologically meaningful, we compare the subgrouping results of cells and their cancer types, and found that the identified subgroups are strongly associated with the recorded cancer types ($\chi^2$ test, $P<0.001$).

\begin{table}
\begin{center}
    \setlength{\tabcolsep}{2.5pt}
    \renewcommand{\arraystretch}{1.0}
\caption{Analysis of CCLE data (drug Paclitaxel) using the proposed approach: identified genes and coefficient estimates for the four subgroups.}
\label{tab:ccle_coef_rks}
\begin{tabular}{lcccc}
\hline
\multicolumn{1}{c}{Gene} & Subgroup 1 & Subgroup 2 & Subgroup 3 & Subgroup 4 \\ \hline
Intercept                & 4.712      & 5.011      & 5.265      & 5.290      \\
RPL35A                   & 0.250      &            &            &            \\
ATG3                     & 0.196      &            &            &            \\
RAMP3                    & 0.148      &            &            &            \\
ISY1                     & 0.134      &            &            &            \\
DNAI2                    & 0.131      &            &            &            \\
AQR                      & 0.103      &            &            &            \\
PLA2G5                   & -0.120     &            &            &            \\
C1QB                     & -0.160     &            &            &            \\
CLTC                     & -0.230     &            &            &            \\
TAF1L                    &            & 0.213      &            & -0.108     \\
KCNJ1                    &            & 0.184      &            &            \\
TAF9                     &            & 0.173      &            & 0.169      \\
OR2L13                   &            & 0.132      &            &            \\
COX10                    &            & 0.105      &            & 0.192      \\
EXOSC3                   &            & -0.104     &            &            \\
OR5V1                    &            & -0.135     &            &            \\
ATP6V0E1                 &            & -0.160     &            &            \\
ATP6V1D                  &            & -0.322     &            & -0.171     \\
EXOSC7                   &            & -0.379     &            &            \\
PPP2R5B                  &            & -0.389     &            &            \\
RSL24D1                  &            &            & 0.293      &            \\
GTF2F1                   &            &            & 0.249      &            \\
HUWE1                    &            &            & 0.141      &            \\
CIITA                    &            &            & 0.124      &            \\
SERPINC1                 &            &            & 0.118      &            \\
RPL13                    &            &            & 0.117      &            \\
RPL10                    &            &            & -0.134     &            \\
UBA2                     &            &            & -0.137     &            \\
FRS2                     &            &            & -0.207     &            \\
COX5A                    &            &            & -0.304     &            \\
POLE3                    &            &            &            & 0.321      \\
KLRD1                    &            &            &            & 0.128      \\ \hline
\end{tabular}
\end{center}
\end{table}

The identified genes and corresponding estimates along with the intercepts for the four subgroups are shown in Table \ref{tab:ccle_coef_rks}. Overall, the four subgroups have significantly different sets of important genes. These genes appear in 44 pathways, of which 11 genes appear in multiple pathways. Table \ref{tab:pathway_rks} lists the identified pathways for each subgroup, from which the differences across the four subgroups are clearly observed. In addition, it is noted that, within some subgroups, multiple identified genes are from the same pathway. For example, in subgroup 2, genes TAF1L and TAF9 participate in basal transcription factors, and in subgroup 3, genes HUWE1 and UBA2 participate in ubiquitin mediated proteolysis. The literature suggests that the identified genes are biologically meaningful. For example, recent studies have revealed that gene UBA2 plays a vital role in gastric cancer cell migration and invasion. It has been found that the level of UBA2 expression significantly increases in colorectal cancer tissues compared to normal tissues. Also, gene UBA2 is strongly associated with survival in colorectal cancer \cite{He2018}. Gene RPL10 takes part in signal transduction, metabolism, cell proliferation, and differentiation \cite{Oh2002}. A high RPL10 protein level has been found during the progression and development of prostate cancer \cite{Altinok2006}. Mutations in the gene RPL10 have been found in patients with autism and modulating disease mechanisms \cite{Chiocchetti2011}.

\begin{table}[]
\caption{Analysis of CCLE data (drug Paclitaxel) using the proposed approach: the identified pathways for the four subgroups. Each cell shows the number of identified genes from the corresponding pathway.}
\label{tab:pathway_rks}
\resizebox{\textwidth}{!}{%
\begin{tabular}{@{}lcccc@{}}
\toprule
\multicolumn{1}{c}{Pathway}                                                                                                                                                                                                                                                                                                                                                                                                     & Subgroup 1                & Subgroup 2               & Subgroup 3               & Subgroup 4               \\ \midrule
\rowcolor[HTML]{EFEFEF} 
Huntingtons Disease                                                                                                                                                                                                                                                                                                                                                                                                             & 2                         &                          & 1                        &                          \\
Spliceosome, Vascular Smooth Muscle Contraction                                                                                                                                                                                                                                                                                                                                                                                 & 2                         &                          &                          &                          \\
\rowcolor[HTML]{EFEFEF} 
Ribosome                                                                                                                                                                                                                                                                                                                                                                                                                        & 1                         &                          & 3                        &                          \\
Complement and Coagulation Cascades                                                                                                                                                                                                                                                                                                                                                                                             & 1                         &                          & 1                        &                          \\
\begin{tabular}[c]{@{}l@{}}\cellcolor[HTML]{EFEFEF} Alpha Linolenic Acid Metabolism, Arachidonic Acid Metabolism, \\ \cellcolor[HTML]{EFEFEF} Endocytosis, Ether Lipid Metabolism, Fc Epsilon RI Signaling Pathway, \\ \cellcolor[HTML]{EFEFEF}Glycerophospholipid Metabolism, GNRH Signaling Pathway, Linoleic Acid Metabolism, \\ \cellcolor[HTML]{EFEFEF}Long Term Depression, Lysosome, MAPK Signaling Pathway, Prion Diseases, \\ \cellcolor[HTML]{EFEFEF} Regulation of Autophagy, Systemic Lupus Erythematosus, VEGF Signaling Pathway\end{tabular} & \cellcolor[HTML]{EFEFEF}1 & \cellcolor[HTML]{EFEFEF} & \cellcolor[HTML]{EFEFEF} & \cellcolor[HTML]{EFEFEF} \\
Oxidative Phosphorylation                                                                                                                                                                                                                                                                                                                                                                                                       &                           & 3                        & 1                        & 2                        \\
\rowcolor[HTML]{EFEFEF} 
Basal Transcription Factors                                                                                                                                                                                                                                                                                                                                                                                                     &                           & 2                        & 1                        & 2                        \\
Epithelial Cell Signaling in Helicobacter Pylori Infection                                                                                                                                                                                                                                                                                                                                                                      &                           & 2                        &                          & 1                        \\
\rowcolor[HTML]{EFEFEF} 
Vibrio Cholerae Infection                                                                                                                                                                                                                                                                                                                                                                                                       &                           & 2                        &                          & 1                        \\
Olfactory Transduction, RNA Degradation                                                                                                                                                                                                                                                                                                                                                                                         &                           & 2                        &                          &                          \\
\rowcolor[HTML]{EFEFEF} 
Porphyrin And Chlorophyll Metabolism                                                                                                                                                                                                                                                                                                                                                                                            &                           & 1                        &                          & 1                        \\
\begin{tabular}[c]{@{}l@{}}Aldosterone Regulated Sodium Reabsorption, \\ Oocyte Meiosis, Oocyte Meiosis, WNT Signaling Pathway\end{tabular}                                                                                                                                                                                                                                                                                     &                           & 1                        &                          &                          \\
\rowcolor[HTML]{EFEFEF} 
Ubiquitin Mediated Proteolysis                                                                                                                                                                                                                                                                                                                                                                                                  &                           &                          & 2                        &                          \\
Antigen Processing and Presentation                                                                                                                                                                                                                                                                                                                                                                                             &                           &                          & 1                        & 1                        \\
\begin{tabular}[c]{@{}l@{}}\cellcolor[HTML]{EFEFEF}Alzheimers Disease, Cardiac Muscle Contraction, \\ \cellcolor[HTML]{EFEFEF} Neurotrophin Signaling Pathway, Parkinsons Disease, Primary Immunodeficiency\end{tabular}   \cellcolor[HTML]{EFEFEF}                                                                                                                                                                                                                                                      & \cellcolor[HTML]{EFEFEF}                          & \cellcolor[HTML]{EFEFEF}                         & \cellcolor[HTML]{EFEFEF}1                        & \cellcolor[HTML]{EFEFEF}                         \\
\begin{tabular}[c]{@{}l@{}}Base Excision Repair, Dna Replication, Graft Versus Host Disease, \\ Natural Killer Cell Mediated Cytotoxicity, Nucleotide Excision Repair,   \\ Nucleotide Excision Repair, Purine Metabolism, Pyrimidine Metabolism\end{tabular}                                                                                                                                                                   &                           &                          &                          & 1                        \\ \bottomrule
\end{tabular}%
}
\end{table}

\begin{figure}
	\centering  
	\subfigure
	{
   	  \includegraphics[width=0.4\textwidth]{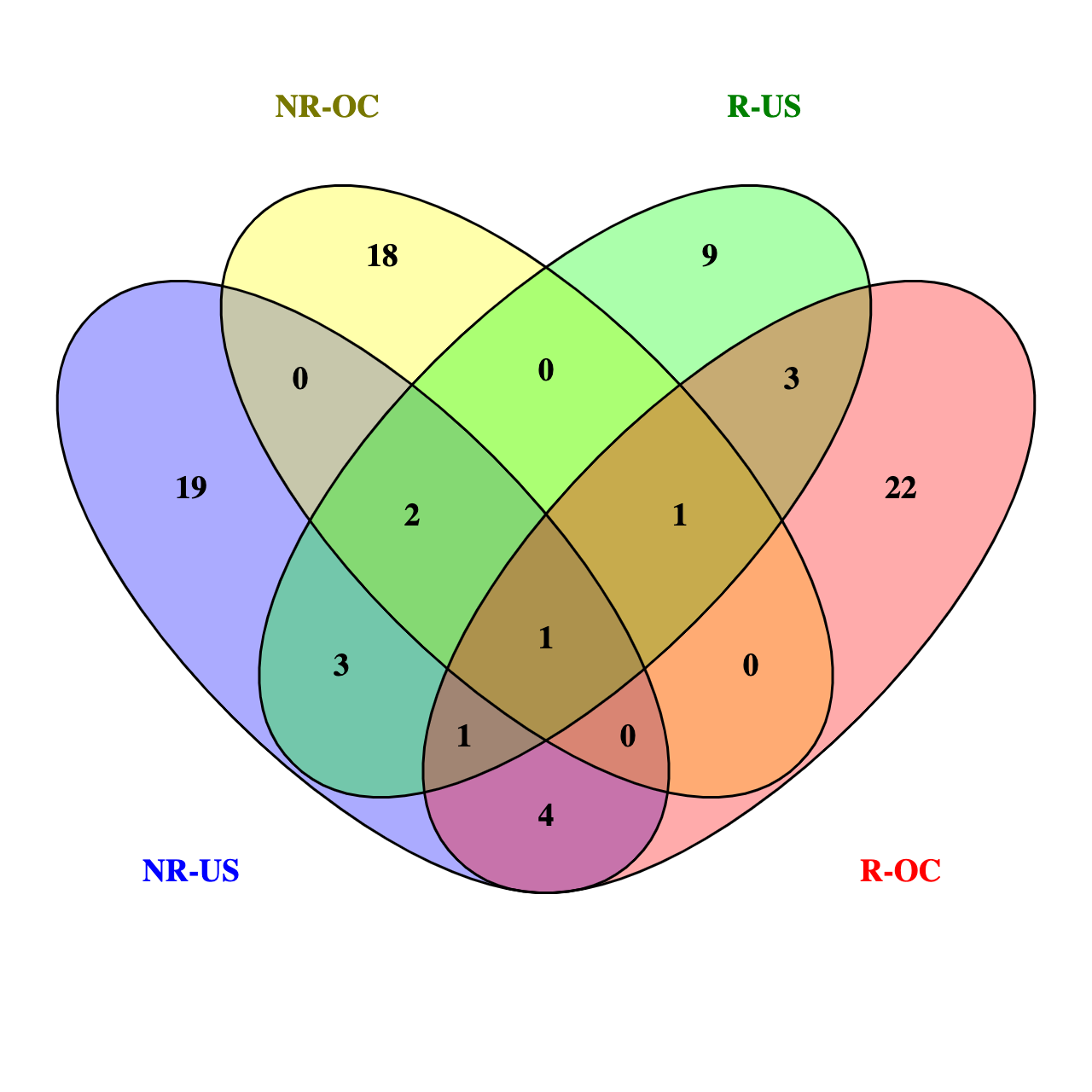}
	}
	\subfigure
	{
	 \includegraphics[width=0.4\textwidth]{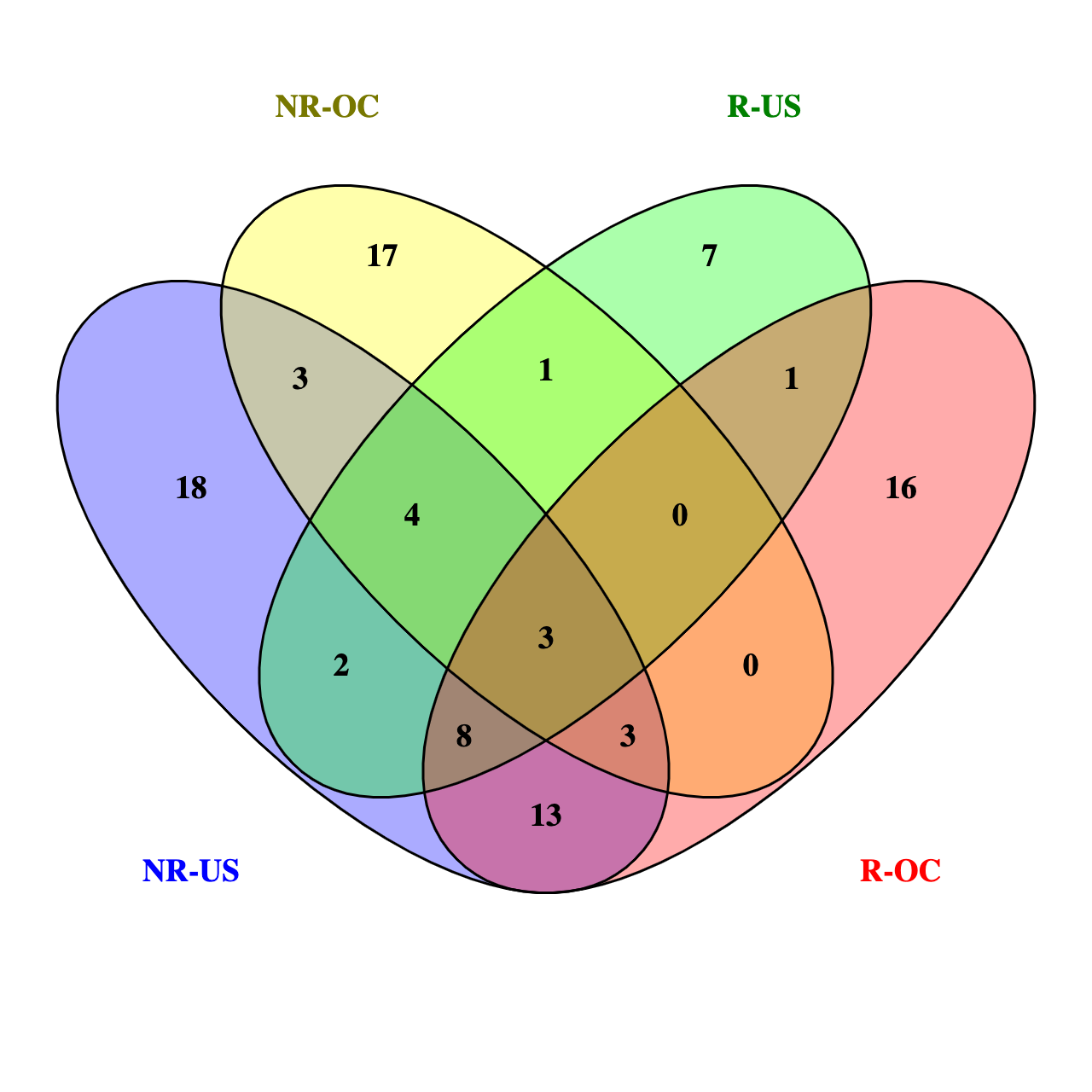} 
	}
	\caption{Analysis of CCLE data (drug Paclitaxel)): comparison of gene identification results. Left: Venn diagram of the identified genes. Right: Venn diagram of the identified pathways. The numbers show the number of overlapping identification. }
	\label{fig:compare}
\end{figure}

Besides the proposed approach, data are also analyzed using the alternatives. We select the number of subgroups by the same BIC criterion as used for the proposed approach. The number of identified subgroups are 5(NR-US), 4(NR-OC), and 3(R-US). We compute the Normalized Mutual Information (NMI) to measure the similarity between the subgrouping results, with range $[0,1]$ and a larger value indicating a higher degree of similarity. The results are provided in Table \ref{tab:nmi} (Appendix \ref{sec:appB}). It can be found that the proposed approach has low similarity with the alternatives. None of the subgroups identified by the alternatives is significantly associated with the recorded subtypes ($\chi^2$ test, $P\in[0.171,0.562]$). The gene identification results are compared in Figure \ref{fig:compare} and Table \ref{tab:compare} (Appendix \ref{sec:appB}). Overall, it is found that different approaches lead to different findings. Details estimation results are provided in Table \ref{tab:ccle_coef_ku}-\ref{tab:ccle_coef_rku} (Appendix \ref{sec:appB}). 

In practical data analysis, it is challenging to objectively assess which set of results is more sensible. To complement the subgrouping and identification results, we evaluate prediction and stability, which may provide support to the analysis results to a certain extent. We first evaluate prediction performance. Specifically, each dataset is randomly split into a training and testing set with sizes 7:3, where the training set is used to obtain the number of subgroups $\hat{K}$, subgroup structure $\hat{\mathcal{G}}$ and subgroup-specific coefficients $\hat{\bm \beta}$, and the testing set is to make prediction. Specifically, for a sample $i$ drawn from the testing set, its subgroup membership is predicted as $\hat{g}_i=\min_{1\leq k\leq K} \rho_{\delta}(y_i-X_i^\top\hat{\beta}_k)$. After assigning the sample to subgroup $\hat{g}_i$, we could make prediction of the response based on linear regression, i.e., $\hat{y}_i=X_i
^\top \hat{\beta}_{\hat{g}_i}$. We use the prediction mean relative error (PMRE) to measure the prediction performance, which is defined as $\sum_{i=1}^n\left|(y
_i-\hat{y}_i)/y_i\right|/n$. 
The process is repeated 100 times. Figure \ref{fig:prediction} (Appendix \ref{sec:appB}) shows the distribution of PMRE for all 24 drugs. It is noted that R-OC has significantly smaller PMRE for all drugs (one-way ANOVA test, $P<0.001$). For example, for the drug Paclitaxel, the median values of PMRE are 0.212(NR-US), 0.208(NR-OC), 0.189(R-US), and 0.164(R-OC). The results demonstrate the consistent and robust performance of R-OC over the alternatives. We also examine the stability of subgrouping. Specifically, 30\% of samples are randomly removed from each dataset. Let $\bm{G} = (G_{il})_{n \times n}$ be the co-existence matrix, where the $(i,l)$th element $G_{il}=1$ if samples $i,l$ belong to the same subgroup, and 0 otherwise. The stability measure $\bm{M}_{sta}$ is defined as $\bm{M}_{sta} = 1/n^2\sum_{i,l=1}^n|\hat{\bm{G}}-\bm{G}^{(T)}|$, where $\hat{\bm{G}}$ and $\bm{G}^{(T)}$ are the co-existence matrices of the estimated and true subgroups, respectively. Here, the true subgroups refer to the identified subgroups by using all samples. A smaller $\bm{M}_{sta}$ indicates a higher stability. For the drug Paclitaxel, the median values of $\bm{M}_{sta}$ over 100 random replicates are 0.411(NR-US), 0.405(NR-OC), 0.393(R-US), and 0.389(R-OC). Similar patterns are observed for other drugs.

It has been recognized in some studies that simulated data may be ``simpler'' than real data. Here we conduct an additional set of simulations based on the CCLE data analyzed above. Specifically, we again take the response Paclitaxel as an example. The observed cancer cell line measurements, overlapping group structure, estimated number of subgroups, and estimated coefficients and intercepts, obtained using R-OC, are used in the simulation. We consider three types of errors described in Section \ref{sec:sim}. The simulation results are summarized in Table \ref{tab:app_simu} (Appendix \ref{sec:appB}). It is observed that the proposed R-OC maintains a relative advantage over the alternatives, which demonstrates the effectiveness of the proposed R-OC.

\section{Conclusion}
Heterogeneity is a hallmark of many complex diseases. With the recent rapid evolution in genomic technology, there has been a surge of heterogeneity analysis based on high-dimensional genomic data. Compared to the heterogeneity in genetic variations, the heterogeneity in the relationship between genetic variations and clinical presentations has been less investigated. In this study, we propose a novel regression-based heterogeneity analysis approach to explicitly deal with the heterogeneous relationship between high-dimensional genetic features and a phenotype of interest. Significantly advancing from the existing literature, 
the proposed approach is robust by adopting a Huber loss function. In addition, we impose a sparse overlapping group lasso penalty to accommodate the high data dimension and screen out noises, while taking into account the overlapping cluster structure of genes. The proposed approach follows the clustering framework, which has appealing performance in high-dimensional data analysis. We develop a two-step iterative algorithm using a similar idea of K-means clustering method. Experimental evaluation on simulated datasets with different settings has demonstrated that the proposed approach outperforms several direct competitors in the identification of subgroups and important genes, and coefficient estimation. In the analysis of CCLE data, findings different from those of the alternatives have been generated, and improved prediction and grouping stability are observed.

This study has focused on the continuous outcome and linear regression model. It will be of interest to extend the proposed analysis to other outcomes/models, e.g., survival outcome and Cox model, binary outcome and Logistic model. The theoretical investigation of the proposed approach is deferred to future research. Last but not the least, it has been recognized that not only genes, but also environments, and gene-environment interactions as well, are associated with the risk and progression of complex diseases. However, most of the existing interaction analysis ignores the possible heterogeneity of the study samples. Heterogeneous interaction analysis is challenged by the ultra high-dimensionality of data and the hierarchy between main and interaction effects. We will investigate it in the followed-up research. 

\section*{Acknowledgments}
We thank the editor and reviewers for their careful review and insightful comments, which have led to a significant improvement of the article. 
This research was supported by the National Natural Science Foundation of China (grants 12171479), and the Fund for building world-class universities (disciplines) of the Renmin University of China. The computer resources were provided by the Public Computing Cloud Platform of the Renmin University of China.

\section*{Data Availability Statement}
The CCLE data analyzed in this study are publicly available at https://sites.broadinstitute.org/ccle/datasets.

\subsection*{Conflict of interest}

The authors declare no potential conflict of interests.
 

\newpage
\appendix

\section{More details in updating $\beta_k$}
\label{app:alg}
Given the group membership $\mathcal{G}=\{g_1,\ldots, g_n\}$, the objective function in (\ref{eq:obj}) is separable with respect to $\beta_k$ ($k=1,\ldots,K$), and thus the original optimization problem can be decomposed into $K$ subproblems: 
\begin{equation}\label{eq:beta}
\hat\beta_k=\mathop{\arg\min}_{\beta\in \mathbb{R}^p}\sum_{\{i|g_i=k\}}\rho_\delta(y_i-X_i^{\top}\beta) + \phi(\beta;\gamma_k,\lambda_k),
\end{equation}
where 
\[\phi(\beta;\gamma_k,\lambda_k)=\lambda_k\inf_{v^{(l)}\in V_l,\sum_{l=1}^L v^{(l)}=\beta}\sum_{l=1}^L \left[ \gamma_k\|v^{(l)}\|_1 + (1-\gamma_k)\omega_l \Vert v^{(l)}\Vert_2\right].\]

A simple way to solve the optimization problem (\ref{eq:beta}) is to convert this problem into the sparse group lasso in an expanded space constructed by explicitly duplicate the covariates belonging to more than one cluster, and adopts any state-of-the-art optimization procedure for sparse group lasso. More precisely, we replace the vector $X_i\in\mathbb{R}^{p}$ by $\tilde{X}_i\in \mathbb{R}^{\sum_{l=1}^L p_l}$ built by concatenating copies of $X_i$ restricted each to a certain cluster $l$, i.e., $\tilde{X}_i=((X_i^{(1)})^{\top},\ldots,\bm (X_i^{(L)})^{\top})^{\top}$, where $X_i^{(l)}$ is the subvector of $X_i$ with entries corresponding to the covariates in cluster $l$. Denote $\tilde{v}^{(l)}=(v_j^{(l)})_{j\in C_l}$ and $\tilde v=((\tilde{v}^{(1)})^\top,\ldots,(\tilde{v}^{(L)})^\top)^\top$. Since $X_i^{\top}\beta=\tilde{X}^{\top}_i\tilde{v}$, we can thus reformulate the above optimization problem as a sparse non-overlapping group lasso problem:   
\begin{equation}\label{eq:obj2}
\hat{\tilde v}=\mathop{\arg\min}_{\tilde v\in \mathbb{R}^{\sum_{l=1}^L p_l}}\sum_{\{i|g_i=k\}}\rho_\delta(y_i-\tilde{X}_i^{\top}\tilde{v})+\lambda_k\left[ \gamma_k\|\tilde v\|_1 + (1-\gamma_k)\sum_{l=1}^L\omega_l \Vert \tilde v^{(l)}\Vert_2\right]. 
\end{equation}
We adopt the Python package 'Groupyr' \cite{Adam2021} with a Huber loss function to optimize (\ref{eq:obj2}). This package is developed by Simon et al \cite{Simon2013} to solve the sparse group lasso problem. Once the solution $\hat{\tilde v}=((\hat{\tilde{v}}^{(1)})^\top,\ldots,(\hat{\tilde{v}}^{(L)})^\top)^\top$ is obtained in the duplicated space, we can reconstruct regression coefficients $\hat{\beta}_k$ by recombining these duplicates.

\newpage
\section{Simulation setting of low-dimensional data}
\label{sec:appB-}
The $p$ genes belong to $i$ clusters, and each cluster contains $j$ genes with 2 genes of overlap between two successive clusters, where $(i,j)$ are dependent on $p$. Specifically, for $p = 10, 20, 32, 50$, $(i,j)$ are  set to be $(2, 6)$, $(3, 8)$, $(5, 8)$ and $(6, 10)$, respectively. For $p = 10$, each subgroup has 3 important genes. In the first subgroup, the set of important genes is $\{1, 2, 4\}$, and their coefficients are $\{2, 1, 0.5\}$; in the second subgroup, the set of important genes is $\{3, 4, 6\}$, and their coefficients are $\{-2, 1, - 0.5\}$. For $p = 20$, each subgroup has 6 important genes. 
In the first subgroup, the set of important genes is $\{1, 2, 6, 8, 10, 11\}$, and their coefficients are $\{2, 1, 0.5,- 1, 1.5, -2\}$; in the second subgroup, the set of important genes is $\{1, 3, 4, 7, 10, 14\}$, and their coefficients are $\{-2, 1, 0.5, -1, 1.5, 2\}$. For $p = 32$ and $50$, in the first subgroup, the set of important genes is $\{1, 8, 10, 13, 16\}$, and their coefficients are $\{2, 1, 0.5, -1, 1.5, -2\}$; in the second subgroup, the set of important genes is $\{1, 4, 7, 10, 14, 19\}$, and their coefficients are $\{-2,1,0.5,-1, 1.5, 2\}$. The $n=300$ samples belong to 2 subgroups with equal size. The other settings are the same as described in Section \ref{sec:sim}.

\newpage

\section{Additional numerical results} \label{sec:appB}

\begin{table}[]\small
\centering
\caption{Simulation scenarios S1-S3 for balanced subgroup structure with $n=500$. Each cell shows the median (median absolute deviation).}
\label{tab:n5bal1}
\resizebox{\textwidth}{!}{%
\begin{tabular}{ccccccc}
\hline
Error                                & Method & ARI                   & TPR                   & FPR                   & MCC                   & RMSE                  \\ \hline
\multicolumn{7}{c}{S1}                                                                                                                                                \\
\multirow{4}{*}{$t(1)$}              & NR-US  & 0.001(0.011)          & 0.117(0.106)          & 0.015(0.082)          & 0.164(0.152)          & 0.421(0.137)          \\
                                     & NR-OC  & 0.008(0.012)          & 0.133(0.205)          & 0.011(0.177)          & 0.215(0.166)          & 0.462(0.180)          \\
                                     & R-US   & 0.487(0.142)          & 0.900(0.248)          & 0.031(0.058)          & 0.667(0.203)          & 0.199(0.054)          \\
                                     & R-OC   & \textbf{0.543(0.146)} & \textbf{1.000(0.246)} & \textbf{0.016(0.016)} & \textbf{0.857(0.179)} & \textbf{0.174(0.053)} \\
                                     &        & \textbf{}             & \textbf{}             & \textbf{}             & \textbf{}             & \textbf{}             \\
\multirow{4}{*}{$0.7N(0,1)+0.3t(1)$} & NR-US  & 0.095(0.100)          & 0.550(0.227)          & 0.135(0.188)          & 0.295(0.037)          & 0.314(0.178)          \\
\hline
\end{tabular}%
}
\end{table}

\begin{table}[]\small
\centering
\caption{Simulation scenarios S4-S6 for balanced subgroup structure with $n=500$. Each cell shows the median (median absolute deviation).}
\label{tab:n5bal2}
\resizebox{\textwidth}{!}{%
\begin{tabular}{ccccccc}
\hline
Error                                & Method & ARI                   & TPR                   & FPR                   & MCC                   & RMSE                  \\ \hline
\multicolumn{7}{c}{S4}                                                                                                                                                \\
\multirow{4}{*}{$t(1)$}              & NR-US  & 0.001(0.012)          & 0.017(0.079)          & 0.003(0.054)          & 0.020(0.077)          & 0.411(0.276)          \\
                                     & NR-OC  & 0.000(0.007)          & 0.067(0.153)          & 0.015(0.102)          & 0.063(0.066)          & 0.465(0.275)          \\
                                     & R-US   & 0.461(0.201)          & 0.850(0.252)          & 0.035(0.066)          & 0.729(0.229)          & 0.201(0.073)          \\
                                     & R-OC   & \textbf{0.520(0.200)} & \textbf{1.000(0.278)} & \textbf{0.038(0.062)} & \textbf{0.803(0.247)} & \textbf{0.182(0.072)} \\
                                     &        &                       &                       &                       &                       &                       \\
\multirow{4}{*}{$0.7N(0,1)+0.3t(1)$} & NR-US  & 0.101(0.050)          & 0.467(0.248)          & 0.274(0.210)          & 0.053(0.089)          & 0.555(0.384)          \\
\hline
\end{tabular}%
}
\end{table}

\begin{table}[]\small
\centering
\caption{Simulation scenarios S1-S3 for unbalanced subgroup structure with $n=300$. Each cell shows the median (median absolute deviation).}
\label{tab:n3unbal1}
\resizebox{\textwidth}{!}{%
\begin{tabular}{ccccccc}
\hline
Error                                & Method & ARI                   & TPR                   & FPR                   & MCC                   & RMSE                  \\ \hline
\multicolumn{7}{c}{S1}                                                                                                                                                \\
\multirow{4}{*}{$t(1)$}              & NR-US  & 0.006(0.015)          & 0.052(0.121)          & \textbf{0.008(0.060)} & 0.091(0.072)          & 0.400(0.159)          \\
                                     & NR-OC  & 0.002(0.020)          & 0.064(0.157)          & 0.018(0.104)          & 0.083(0.101)          & 0.404(0.101)          \\
                                     & R-US   & 0.107(0.109)          & 0.659(0.171)          & 0.137(0.077)          & 0.370(0.137)          & 0.252(0.040)          \\
                                     & R-OC   & \textbf{0.460(0.128)} & \textbf{0.887(0.175)} & 0.023(0.034)          & \textbf{0.837(0.143)} & \textbf{0.200(0.034)} \\
                                     &        & \textbf{}             & \textbf{}             & \textbf{}             & \textbf{}             &                       \\
\multirow{4}{*}{$0.7N(0,1)+0.3t(1)$} & NR-US  & 0.041(0.078)          & 0.300(0.203)          & 0.096(0.092)          & 0.158(0.115)          & 0.409(0.115)          \\
\hline
\end{tabular}%
}
\end{table}

\begin{table}[]\small
\centering
\caption{Simulation scenarios S4-S6 for unbalanced subgroup structure with $n=300$. Each cell shows the median (median absolute deviation).}
\label{tab:n3unbal2}
\resizebox{\textwidth}{!}{%
\begin{tabular}{ccccccc}
\hline
Error                                & Method & ARI                   & TPR                   & FPR                   & MCC                   & RMSE                  \\ \hline
\multicolumn{7}{c}{S4}                                                                                                                                                \\
\multirow{4}{*}{$t(1)$}              & NR-US  & -0.001(0.007)         & 0.067(0.069)          & 0.003(0.057)          & 0.051(0.081)          & 0.418(0.110)          \\
                                     & NR-OC  & 0.001(0.015)          & 0.050(0.093)          & 0.018(0.053)          & 0.003(0.084)          & 0.413(0.126)          \\
                                     & R-US   & 0.005(0.039)          & 0.283(0.137)          & 0.124(0.091)          & 0.110(0.084)          & 0.372(0.033)          \\
                                     & R-OC   & \textbf{0.308(0.058)} & \textbf{0.664(0.147)} & \textbf{0.095(0.066)} & \textbf{0.404(0.123)} & \textbf{0.176(0.032)} \\
                                     &        & \textbf{}             & \textbf{}             & \textbf{}             & \textbf{}             &                       \\
\multirow{4}{*}{$0.7N(0,1)+0.3t(1)$} & NR-US  & 0.070(0.051)          & 0.220(0.192)          & 0.058(0.091)          & 0.152(0.094)          & 0.422(0.277)          \\
\hline
\end{tabular}%
}
\end{table}

\begin{table}[]\small
\centering
\caption{Simulation scenarios S1-S3 for unbalanced subgroup structure with $n=500$. Each cell shows the median (median absolute deviation).}
\label{tab:n5unbal1}
\resizebox{\textwidth}{!}{%
\begin{tabular}{ccccccc}
\hline
Error                                & Method & ARI                   & TPR                   & FPR                   & MCC                   & RMSE                  \\ \hline
\multicolumn{7}{c}{S1}                                                                                                                                                \\
\multirow{4}{*}{$t(1)$}              & NR-US  & 0.002(0.01)           & 0.110(0.118)          & 0.021(0.092)          & 0.148(0.158)          & 0.493(0.158)          \\
                                     & NR-OC  & 0.005(0.018)          & 0.154(0.217)          & 0.019(0.184)          & 0.213(0.165)          & 0.527(0.198)          \\
                                     & R-US   & 0.476(0.141)          & 0.892(0.215)          & 0.046(0.065)          & 0.676(0.218)          & 0.232(0.056)          \\
                                     & R-OC   & \textbf{0.530(0.118)} & \textbf{0.975(0.191)} & \textbf{0.019(0.017)} & \textbf{0.859(0.157)} & \textbf{0.207(0.04)}  \\
                                     &        & \textbf{}             & \textbf{}             & \textbf{}             & \textbf{}             &                       \\
\multirow{4}{*}{$0.7N(0,1)+0.3t(1)$} & NR-US  & 0.100(0.106)          & 0.520(0.254)          & 0.158(0.156)          & 0.257(0.085)          & 0.408(0.152)          \\
\hline
\end{tabular}%
}
\end{table}

\begin{table}[]\small
\centering
\caption{Simulation scenarios S4-S6 for unbalanced subgroup structure with $n=500$. Each cell shows the median (median absolute deviation).}
\label{tab:n5unbal2}
\resizebox{\textwidth}{!}{%
\begin{tabular}{ccccccc}
\hline
Error                                & Method & ARI                   & TPR                   & FPR                   & MCC                   & RMSE                  \\ \hline
\multicolumn{7}{c}{S4}                                                                                                                                                \\
\multirow{4}{*}{$t(1)$}              & NR-US  & 0.002(0.016)          & 0.033(0.090)          & 0.003(0.067)          & 0.006(0.093)          & 0.388(0.235)          \\
                                     & NR-OC  & 0.000(0.021)          & 0.033(0.189)          & 0.007(0.110)          & 0.014(0.070)          & 0.442(0.300)          \\
                                     & R-US   & 0.433(0.068)          & 0.800(0.132)          & 0.031(0.045)          & 0.680(0.130)          & 0.210(0.032)          \\
                                     & R-OC   & \textbf{0.442(0.082)} & \textbf{0.850(0.139)} & \textbf{0.027(0.045)} & \textbf{0.695(0.138)} & \textbf{0.202(0.035)} \\
                                     &        & \textbf{}             & \textbf{}             & \textbf{}             & \textbf{}             &                       \\
\multirow{4}{*}{$0.7N(0,1)+0.3t(1)$} & NR-US  & 0.129(0.154)          & 0.600(0.230)          & 0.079(0.144)          & 0.270(0.230)          & 0.417(0.181)          \\
\hline
\end{tabular}%
}
\end{table}

\begin{table}[]\small
\centering
\caption{Simulation results: the sample mean, median, standard error (s.e.) of $\hat{K}$ and the percentage of $\hat{K}$ equaling the true number of subgroups ($K=3$) by the proposed R-OC.}
\label{tab:k3select}
\resizebox{\textwidth}{!}{%
\begin{tabular}{cccccccccc}
\hline
\multirow{2}{*}{Error} & \multicolumn{4}{c}{$n=300$}         &  & \multicolumn{4}{c}{$n=500$}         \\ \cline{2-5} \cline{7-10} 
                       & Mean & Median & s.e. & Percentage &  & Mean & Median & s.e. & Percentage \\ \hline
$t(1)$                 & 2.58 & 3.00   & 0.55 & 0.71       &  & 2.86 & 3.00   & 0.43 & 0.89       \\
$0.7N(0,1)+0.3t(1)$    & 3.28 & 3.00   & 0.42 & 0.88       &  & 3.06 & 3.00   & 0.34 & 0.93       \\
$N(0,1)$               & 3.45 & 3.00   & 0.59 & 0.70       &  & 3.30 & 3.00   & 0.37 & 0.80       \\ \hline
\end{tabular}%
}
\end{table}

\begin{table}[]\small
\centering
\caption{Simulation scenario S1 for $K=3$ subgroups with $n=300$. Each cell shows the median (median absolute deviation).}
\label{tab:k3}
\resizebox{\textwidth}{!}{%
\begin{tabular}{ccccccc}
\hline
Error                                & Method & ARI                   & TPR                   & FPR                   & MCC                   & RMSE                  \\ \hline
\multirow{4}{*}{$t(1)$}              & NR-US  & 0.000(0.011)          & 0.167(0.100)          & 0.091(0.068)          & 0.068(0.065)          & 0.532(0.011)          \\
                                     & NR-OC  & 0.001(0.009)          & 0.111(0.128)          & 0.063(0.12)           & 0.051(0.08)           & 0.445(0.013)          \\
                                     & R-US   & 0.005(0.082)          & 0.244(0.139)          & 0.107(0.08)           & 0.112(0.118)          & 0.397(0.042)          \\
                                     & R-OC   & \textbf{0.459(0.142)} & \textbf{0.669(0.170)} & \textbf{0.053(0.029)} & \textbf{0.543(0.179)} & \textbf{0.296(0.091)} \\                                                                                                                                             \\
\multirow{4}{*}{$0.7N(0,1)+0.3t(1)$} & NR-US  & 0.084(0.105)          & 0.300(0.162)          & 0.082(0.106)          & 0.193(0.094)          & 0.416(0.112)          \\
                                     & NR-OC  & 0.096(0.132)          & 0.344(0.262)          & \textbf{0.073(0.193)} & 0.245(0.159)          & 0.403(0.157)          \\
\hline
\end{tabular}%
}
\end{table}

\begin{table}[]\small
\centering
\caption{Simulation for low-dimensional data under balanced subgroup structure with $K=2$, $n=300$ and $\varepsilon_i \sim N(0,1)$. Each cell shows the median (median absolute deviation).}
\label{tab:penalized_fusion_n}
\resizebox{\textwidth}{!}{%
\begin{tabular}{cclllll}
\hline
$p$                   & Method           & \multicolumn{1}{c}{ARI} & \multicolumn{1}{c}{TPR} & \multicolumn{1}{c}{FPR} & \multicolumn{1}{c}{MCC} & \multicolumn{1}{c}{RMSE} \\ \hline
\multirow{5}{*}{10}  & Fusion & 0.597(0.083)            & \textbf{1.000(0.000)}   & 1.000(0.000)            & 0.000(0.000)            & 0.269(0.032)             \\ 
                    & NR-US            & 0.858(0.039)            & \textbf{1.000(0.000)}   & \textbf{0.000(0.024)}   & \textbf{1.000(0.035)}   & 0.182(0.019)             \\
                    & NR-OC            & \textbf{0.888(0.039)}   & \textbf{1.000(0.000)}   & \textbf{0.000(0.032)}   & \textbf{1.000(0.047)}   & 0.182(0.018)             \\
                    & R-US             & 0.862(0.038)            & \textbf{1.000(0.000)}   & \textbf{0.000(0.013)}   & \textbf{1.000(0.019)}   & 0.182(0.018)             \\
                    & R-OC             & 0.883(0.036)            & \textbf{1.000(0.000)}   & \textbf{0.000(0.013)}   & \textbf{1.000(0.019)}   & \textbf{0.180(0.018)}    \\ \hline
\multirow{5}{*}{20} 
                    & Fusion & 0.532(0.096)            & \textbf{1.000(0.000)}   & 1.000(0.000)            & 0.000(0.000)            & 0.260(0.031)             \\ 
                    & NR-US            & 0.864(0.039)            & \textbf{1.000(0.000)}   & \textbf{0.000(0.017)}   & \textbf{1.000(0.026)}   & 0.191(0.028)             \\
\hline
\end{tabular}%
}
\end{table}

\begin{table}[]\small
\centering
\caption{Simulation for low-dimensional data under balanced subgroup structure with $K=2$, $n=300$ and $\varepsilon_i \sim 0.7N(0, 1) + 0.3t(1)$. Each cell shows the median (median absolute deviation).}
\label{tab:penalized_fusion_nt}
\resizebox{\textwidth}{!}{%
\begin{tabular}{cclllll}
\hline
p                   & Method           & \multicolumn{1}{c}{ARI} & \multicolumn{1}{c}{TPR} & \multicolumn{1}{c}{FPR} & \multicolumn{1}{c}{MCC} & \multicolumn{1}{c}{RMSE} \\ \hline
\multirow{5}{*}{10} 
                    & Fusion & 0.244(0.051)            & \textbf{1.000(0.000)}   & 1.000(0.000)            & 0.000(0.000)            & 0.656(0.077)             \\ 
                    & NR-US            & 0.202(0.159)            & 0.783(0.242)            & 0.329(0.204)            & 0.302(0.175)            & 0.547(0.153)             \\
                    & NR-OC            & 0.204(0.149)            & 0.917(0.231)            & 0.300(0.194)            & 0.458(0.210)            & 0.536(0.105)             \\
                    & R-US             & 0.671(0.027)            & \textbf{1.000(0.075)}   & 0.071(0.055)            & 0.872(0.092)            & 0.354(0.055)             \\
                    & R-OC             & \textbf{0.751(0.034)}   & \textbf{1.000(0.082)}   & \textbf{0.036(0.064)}   & \textbf{0.897(0.102)}   & \textbf{0.262(0.047)}    \\
                    \hline
\multirow{5}{*}{20} 
                    & Fusion & 0.200(0.067)            & \textbf{1.000(0.000)}   & 1.000(0.000)            & 0.000(0.000)            & 0.566(0.020)             \\ 
                    & NR-US            & 0.202(0.173)            & 0.725(0.287)            & 0.350(0.248)            & 0.305(0.222)            & 0.531(0.127)             \\
\hline
\end{tabular}%
}
\end{table}

\begin{table}[]\small
\centering
\caption{Simulation for low-dimensional data under balanced subgroup structure with $K=2$, $n=300$ and $\varepsilon_i\sim t(1)$. Each cell shows the median (median absolute deviation).}
\label{tab:penalized_fusion_t}
\resizebox{\textwidth}{!}{%
\begin{tabular}{cclllll}
\hline
p                   & Method           & \multicolumn{1}{c}{ARI} & \multicolumn{1}{c}{TPR} & \multicolumn{1}{c}{FPR} & \multicolumn{1}{c}{MCC} & \multicolumn{1}{c}{RMSE} \\ \hline
\multirow{5}{*}{10} 
                    & Fusion & 0.006(0.079)            & \textbf{1.000(0.000)}   & 1.000(0.000)            & 0.000(0.000)            & 0.867(0.111)             \\ 
                    & NR-US            & 0.001(0.075)            & 0.500(0.213)            & 0.338(0.159)            & 0.100(0.153)            & 0.776(0.165)             \\
                    & NR-OC            & 0.002(0.046)            & 0.502(0.257)            & 0.326(0.189)            & 0.103(0.109)            & 0.699(0.163)             \\
                    & R-US             & 0.316(0.047)            & \textbf{1.000(0.120)}   & \textbf{0.143(0.106)}   & \textbf{0.762(0.123)}   & 0.426(0.045)             \\
                    & R-OC             & \textbf{0.412(0.045)}   & \textbf{1.000(0.101)}   & \textbf{0.143(0.098)}   & \textbf{0.762(0.106)}   & \textbf{0.334(0.036)}    \\
                   \hline
\multirow{5}{*}{20} 
                    & Fusion & 0.039(0.023)            & \textbf{1.000(0.000)}   & 1.000(0.000)            & 0.000(0.000)            & 0.868(0.064)             \\
                    & NR-US            & 0.001(0.056)            & 0.475(0.233)            & 0.314(0.188)            & 0.107(0.119)            & 0.719(0.116)             \\
\hline
\end{tabular}%
}
\end{table}

\begin{table}[]
\centering
\caption{Analysis of CCLE data (drug Paclitaxel): comparison of subgrouping results. Each cell shows the value of NMI.}
\label{tab:nmi}
\begin{tabular}{lcccc}
\hline
\multicolumn{1}{c}{} & NR-US & NR-OC & R-US  & R-OC  \\ \hline
NR-US                & 1.000 & 0.084 & 0.105 & 0.035 \\
NR-OC                &       & 1.000 & 0.047 & 0.092 \\
R-UC                 &       &       & 1.000 & 0.082 \\
R-OC                 &       &       &       & 1.000 \\ \hline
\end{tabular}
\end{table}

\begin{table}[]
\centering
\caption{Analysis of CCLE data (drug Paclitaxel): result comparison of different methods. \#Subgroups denotes the number of subgroups; \#Genes denotes the total number of selected genes; \#Pathways denotes the number of selected pathways.}
\label{tab:compare}
\begin{tabular}{cccc}
\hline
Method & \#Subgroups & \#Genes & \#Pathways \\ \hline
NR-US  & 5           & 30      & 54         \\
NR-OC  & 4           & 22      & 31         \\
R-US   & 3           & 20      & 26         \\
R-OC   & 4           & 32      & 44         \\ \hline
\end{tabular}
\end{table}

\begin{table}
\centering
\caption{Analysis of CCLE data (drug Paclitaxel) using NR-US: identified genes and coefficient estimates for the five subgroups.}
\label{tab:ccle_coef_ku}
\begin{tabular}{lccccc}
\hline
\multicolumn{1}{c}{Gene} & Subgroup 1 & Subgroup 2 & Subgroup 3 & Subgroup 4 & Subgroup 5 \\ \hline
Intercept                & 4.711      & 5.141      & 5.261      & 4.387      & 5.091      \\
UBE2J2                   & 0.228      & 0.136      & 0.127      &            &            \\
NOD1                     & 0.152      &            &            &            &            \\
IFNA10                   & -0.218     &            &            &            &            \\
TAF2                     &            & 0.183      &            &            &            \\
OR10C1                   &            & 0.181      &            &            &            \\
POLE3                    &            & 0.169      &            & -0.221     &            \\
\hline
\end{tabular}
\end{table}

\begin{table}[]
\centering
\caption{Analysis of CCLE data (drug Paclitaxel) using NR-OC: identified genes and coefficient estimates for the four subgroups.}
\label{tab:ccle_coef_ks}
\begin{tabular}{lcccc}
\hline
\multicolumn{1}{c}{Gene} & Subgroup 1 & Subgroup 2 & Subgroup 3 & Subgroup 4 \\ \hline
Intercept                & 4.340      & 5.200      & 5.063      & 5.390      \\
PSMB1                    & 0.306      &            &            &            \\
ASMT                     & 0.244      &            &            &            \\
SNRNP200                 & -0.205     &            &            &            \\
RPA2                     & -0.302     &            &            &            \\
CNOT10                   &            & 0.265      &            &            \\
OR2L13                   &            & 0.193      &            &            \\
BUB1B                    &            & 0.186      &            &            \\
IL12RB1                  &            & 0.180      &            &            \\
BUB3                     &            & 0.156      &            &            \\
PAPOLG                   &            & 0.101      &            &            \\
TACR3                    &            & -0.113     &            &            \\
CYP11B1                  &            & -0.293     &            &            \\
RAD51                    &            &            & 0.144      &            \\
STK11                    &            &            & 0.111      &            \\
INPP5B                   &            &            & 0.104      &            \\
RBM22                    &            &            & -0.228     &            \\
WASL                     &            &            & -0.293     &            \\
CLDN19                   &            &            &            & 0.291      \\
HNRNPA1                  &            &            &            & 0.174      \\
AQR                      &            &            &            & 0.119      \\
RPL8                     &            &            &            & -0.112     \\
SNW1                     &            &            &            & -0.329     \\ \hline
\end{tabular}
\end{table}

\begin{table}[]
\centering
\caption{Analysis of CCLE data (drug Paclitaxel) using R-US: identified genes and coefficient estimates for the three subgroups.}
\label{tab:ccle_coef_rku}
\begin{tabular}{@{}lccc@{}}
\toprule
\multicolumn{1}{c}{Gene} & Subgroup 1 & Subgroup 2 & Subgroup 3 \\ \midrule
Intercept                & 5.054      & 5.040      & 5.236      \\
DAXX                     & 0.316      &            &            \\
PTTG2                    & 0.193      &            &            \\
NTSR2                    & 0.154      &            &            \\
TAF2                     & 0.137      &            &            \\
HK3                      & -0.166     &            &            \\
XCR1                     & -0.188     &            &            \\
ADRA1D                   & -0.238     &            &            \\
FRS2                     & -0.343     & -0.182     &            \\
RSL24D1                  &            & 0.338      &            \\
NCBP1                    &            & 0.155      &            \\
GGA1                     &            & 0.141      &            \\
AQR                      &            & 0.108      &            \\
RBM22                    &            &            & 0.204      \\
UBE2J2                   &            &            & 0.190      \\
SF3A3                    &            &            & 0.164      \\
RAMP3                    &            &            & 0.146      \\
RPL35A                   &            &            & 0.140      \\
OR2L13                   &            &            & 0.122      \\
BUB1B                    &            &            & 0.104      \\
UPP2                     &            &            & -0.118     \\ \bottomrule
\end{tabular}
\end{table}

\begin{table}[]\small
\centering
\caption{Simulation based on the CCLE data (drug Paclitaxel). Each cell shows the median (median absolute deviation).}
\label{tab:app_simu}
\resizebox{\textwidth}{!}{%
\begin{tabular}{cclllll}
\hline
Error                                & Method & \multicolumn{1}{c}{ARI} & \multicolumn{1}{c}{TPR} & \multicolumn{1}{c}{FPR} & \multicolumn{1}{c}{MCC} & \multicolumn{1}{c}{RMSE} \\ \hline
\multirow{4}{*}{$t(1)$}              & NR-US  & 0.018(0.022)            & 0.275(0.148)            & 0.155(0.043)            & 0.125(0.071)            & 0.609(0.041)             \\
                                     & NR-OC  & 0.011(0.025)            & 0.283(0.159)            & \textbf{0.125(0.046)}   & 0.131(0.068)            & 0.608(0.056)             \\
                                     & R-US   & 0.031(0.078)            & 0.486(0.151)            & 0.132(0.034)            & 0.156(0.061)            & 0.572(0.058)             \\
                                     & R-OC   & \textbf{0.241(0.095)}   & \textbf{0.525(0.155)}   & \textbf{0.125(0.041)}   & \textbf{0.219(0.101)}   & \textbf{0.427(0.083)}    \\ \hline
\multirow{4}{*}{$0.7N(0,1)+0.3t(1)$} & NR-US  & 0.137(0.079)            & 0.315(0.176)            & \textbf{0.071(0.032)}   & 0.214(0.119)            & 0.401(0.053)             \\
                                     & NR-OC  & 0.165(0.077)            & 0.354(0.185)            & 0.076(0.023)            & 0.244(0.111)            & 0.404(0.042)             \\
                                     & R-US   & 0.173(0.101)            & 0.678(0.131)            & 0.084(0.022)            & 0.409(0.096)            & 0.332(0.056)             \\
\hline
\end{tabular}%
}
\end{table}

\clearpage
\begin{figure}[H]
  \centering
  \includegraphics[width=1\textwidth]{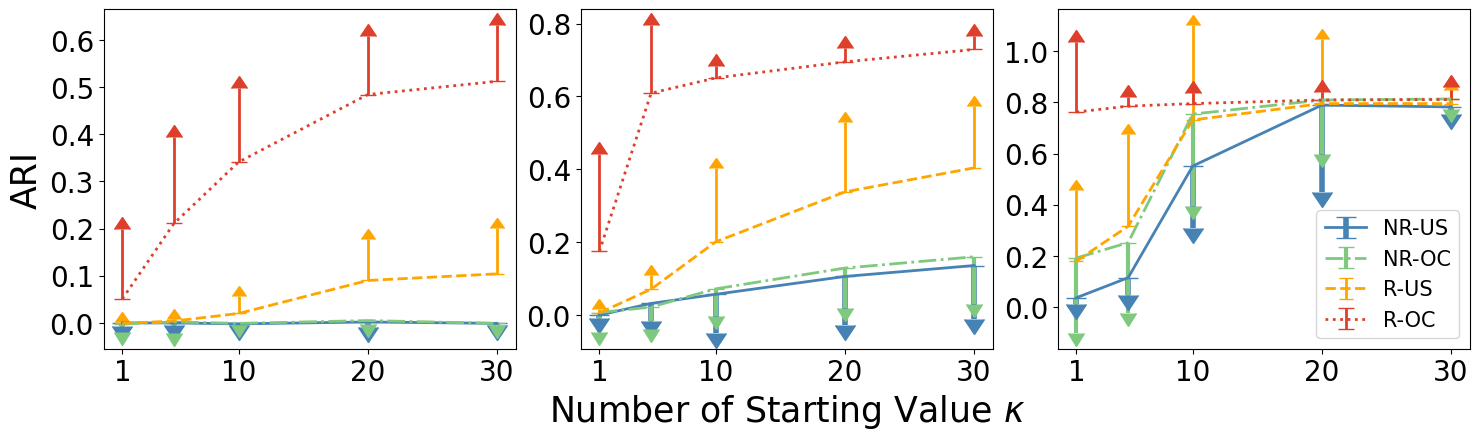}
  \caption{Simulation results: ARI as a function of the number of starting values $\kappa$ based on 100 replicates under scenario S1, balanced structure $n=300$. Left: $\varepsilon_i\sim t(1)$. Middle: $\varepsilon_i\sim 0.7N(0,1)+0.3t(1)$. Right: 
$\varepsilon_i\sim N(0,1)$.  For clarity, only the half (upper or lower) error bars are displayed. }
  \label{fig:init}
\end{figure}

\begin{figure}[H]
  \centering
  \includegraphics[width=0.7\textwidth]{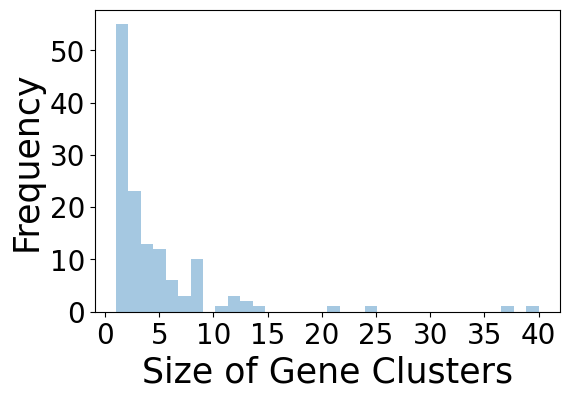}
  \caption{Analysis of CCLE data: histogram of sizes of gene clusters.}
  \label{fig:group_distr}
\end{figure}

\begin{figure}[H]
  \centering
  \includegraphics[width=0.7\textwidth]{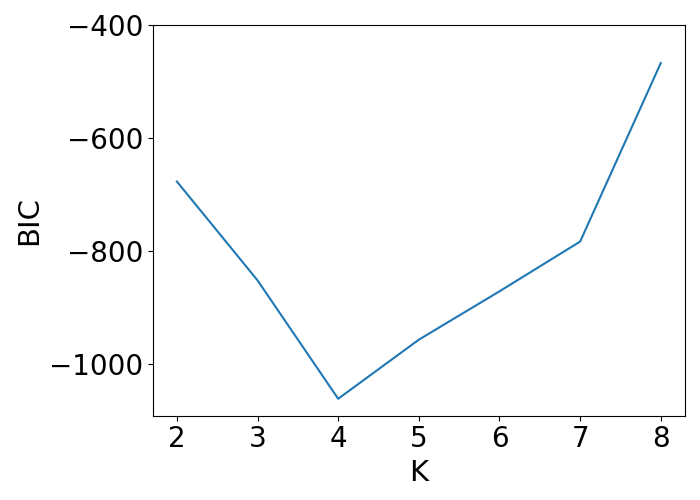}
  \caption{Analysis of CCLE data (drug Paclitaxel) using the proposed approach: BIC as a function of the number of subgroups $K$.}
  \label{fig: bic}
\end{figure}

\begin{figure}[H]
  \centering
  \includegraphics[width=0.9\textwidth]{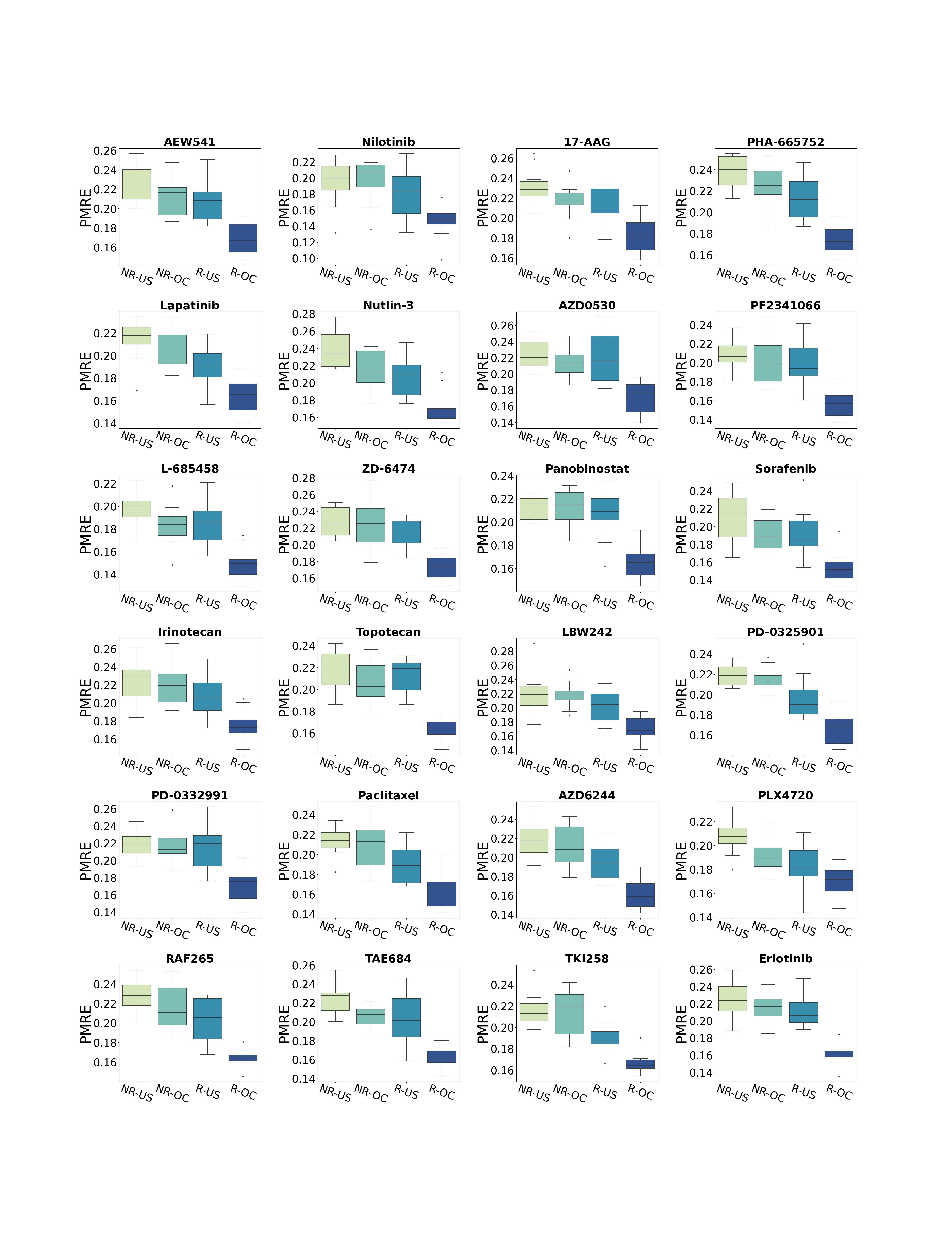}
  \caption{Analysis of CCLE data: distribution of the PMRE over 100 repetitions for the 24 drugs.}
 \label{fig:prediction}
\end{figure}

\end{document}